\documentclass[11pt]{article}

\usepackage{graphicx}
\usepackage{dcolumn}
\usepackage{bm}

\usepackage{amsfonts,amsmath,amssymb,mathrsfs,amsthm}
\usepackage{epsfig,epstopdf,graphicx,graphics}
\usepackage{color}
\usepackage{float}
\usepackage{ifpdf}
\usepackage[colorlinks=true]{hyperref}
\usepackage{braket}
\usepackage{enumerate}
\usepackage{verbatim}
\usepackage{algorithm}
\usepackage{algorithmic}
\usepackage[margin=1in]{geometry}
\usepackage[hang,small,bf]{caption}
\usepackage{tikz}
\usetikzlibrary{backgrounds,shadows.blur,fit,decorations.pathreplacing,shapes}
\usepackage{adjustbox}

\newtheorem{theorem}{Theorem}

\newtheorem{assumption}{Assumption}
\newtheorem{corollary}{Corollary}
\newtheorem{definition}{Definition}
\newtheorem{example}{Example}
\newtheorem{lemma}{Lemma}
\newtheorem{problem}{Problem}
\newtheorem{proposition}{Proposition}
\newtheorem{remark}{Remark}
\newcommand{\bthm}{\begin{theorem}}
	\newcommand{\ethm}{\end{theorem}}
\newcommand{\blem}{\begin{lemma}}
	\newcommand{\elem}{\end{lemma}}
\newcommand{\bex}{\begin{example}}
	\newcommand{\eex}{\end{example}}
\newcommand{\bprop}{\begin{proposition}}
	\newcommand{\eprop}{\end{proposition}}
\newcommand{\bplm}{\begin{problem}}
	\newcommand{\eplm}{\end{problem}}
\newcommand{\bmrk}{\begin{remark}}
	\newcommand{\emrk}{\end{remark}}
\newcommand{\bdfn}{\begin{definition}}
	\newcommand{\edfn}{\end{definition}}
\newcommand{\bcor}{\begin{corollary}}
	\newcommand{\ecor}{\end{corollary}}
\newcommand{\E}{\mathrm{E}}

\newcommand{\beq}{\begin{equation}}
\newcommand{\eeq}{\end{equation}}
\newcommand{\beqm}{\begin{equation*}}
\newcommand{\eeqm}{\end{equation*}}
\newcommand{\beqn}{\begin{eqnarray}}
\newcommand{\eeqn}{\end{eqnarray}}
\newcommand{\beqnm}{\begin{eqnarray*}}
	\newcommand{\eeqnm}{\end{eqnarray*}}
\newcommand{\bea}{\begin{align}}
\newcommand{\eea}{\end{align}}
\newcommand{\beam}{\begin{align*}}
\newcommand{\eeam}{\end{align*}}
\newcommand{\bei}{\begin{itemize}}
	\newcommand{\eei}{\end{itemize}}
\newcommand{\bed}{\begin{description}}
	\newcommand{\eed}{\end{description}}
\newcommand{\bee}{\begin{enumerate}}
	\newcommand{\eee}{\end{enumerate}}
\newcommand{\bey}{\begin{array}}
	\newcommand{\eey}{\end{array}}
\newcommand{\beb}{\begin{thebibliography}{99}}
	\newcommand{\eeb}{\end{thebibliography}}

\newcommand{\ii}{{\rm i}}

\def\r{\rangle}

\begin{document}


\title{Quantum tensor singular value decomposition with applications to recommendation systems}

\author{Xiaoqiang Wang\thanks{xiaoqiang.wang@connect.polyu.hk}, Lejia Gu\thanks{le-jia.gu@connect.polyu.hk}, Joseph Heung-wing Lee\thanks{joseph.lee@polyu.edu.hk}, and Guofeng Zhang\thanks{Guofeng.Zhang@polyu.edu.hk} \\Department of Applied Mathematics, The Hong Kong Polytechnic University, Hong Kong}

\date{\today}
             
\maketitle             

\begin{abstract}
In this paper, we present a quantum singular value decomposition algorithm for third-order tensors inspired by the classical algorithm of tensor singular value decomposition (t-svd) and then extend it to order-$p$ tensors. It can be proved that the quantum version of the t-svd for a third-order tensor $\mathcal{A} \in \mathbb{R}^{N\times N \times N}$ achieves the complexity of $\mathcal{O}(N{\rm polylog}(N))$, an exponential speedup compared with its classical counterpart. As an application, we propose a quantum algorithm for recommendation systems which incorporates the contextual situation of users to the personalized recommendation.
We provide recommendations varying with contexts by measuring the output quantum state corresponding to an approximation of this user's preferences. This algorithm runs in expected time $\mathcal{O}(N{\rm polylog}(N){\rm poly}(k)),$ if every frontal slice of the preference tensor has a good rank-$k$ approximation. At last, we provide a quantum algorithm for tensor completion based on a different truncation method which is tested to have a good performance in dynamic video completion.

\end{abstract}

\maketitle


\section{Introduction}
Tensor refers to a multi-dimensional array of objects. The order of a tensor is the number of modes. For example, $\mathcal{A} \in \mathbb{R}^{N_1 \times N_2 \times N_3}$ is a third-order tensor of real numbers with dimension $N_i$ for mode $i$, $i = 1, 2, 3$, respectively. Due to their flexibility of representing data, tensors have versatile applications in many areas such as image deblurring, video recovery, denoising,  data completion, multi-partite quantum systems, networks and machine learning \cite{t-svd_three, exact_tensor_completion, Novel, 5D,Tensor_factorization_for_low-rank, KB09, Qi, Qi2, rendle2010pairwise, rendle2009learning, HQZ16, Z17, QZB17, QZN18, PMW19, ZNZ19, WHZ19, MWT20}. Some of these practical problems are addressed by different ways of tensor decomposition, including, but not limited to, CANDECOMP/PARAFAC (CP) \cite{CP}, TUCKER \cite{Tucker}, higher-order singular value decomposition (HOSVD) \cite{HOSVD, GWZ19}, Tensor-train decomposition (TT) \cite{oseledets2011tensor} and tensor singular value decomposition (t-svd) \cite{t-svd_three, Novel, MWT20}.

Plenty of research has been carried out on t-svd recently. The concept of t-svd was first proposed by Kilmer and Martin \cite{t-svd_three} for third-order tensors. Later, Martin et al. \cite{order-p} later extended it to higher-order tensors. The t-svd algorithm is superior to TUCKER and CP decompositions in the sense that it extends the familiar matrix svd strategy to tensors efficiently thus avoiding the loss of information inherent in flattening tensors used in TUCKER and CP decompositions. One can obtain a t-svd by computing matrix svd in the Fourier domain, and it also allows other matrix factorization techniques like QR decomposition to be extended to tensors easily in the similar way.

In this paper, we propose a quantum version of t-svd. An important step in a (classical) t-svd algorithm is performing Fast Fourier Transform (FFT) along the third mode of a tensor $\mathcal{A} \in \mathbb{R}^{N_1 \times N_2 \times N_3}$, obtaining $\hat{\mathcal{A}}$ with computational complexity $\mathcal{O}(N_3{\rm log}N_3)$ for each tube $\mathcal{A}(i,j,:),$ $i=0,\cdots, N_1-1, j=0,\cdots, N_2-1$. In the quantum t-svd algorithm to be proposed, this procedure is accelerated by the quantum Fourier transform (QFT) \cite{Nielsen} whose complexity is only $\mathcal{O}(({\rm log}N_3)^2)$. Moreover, due to quantum superposition, the QFT can be performed on the third register of the state $\ket{\mathcal{A}}$, which is equivalent to performing the FFT for all tubes of $\mathcal{A}$ parallelly, so the total complexity of this step is still $\mathcal{O}(({\rm log}N_3)^2)$.

After performing the QFT, in order to further accelerate the second step in the classical t-svd algorithm (the matrix svd), we apply the quantum singular value estimation (QSVE) algorithm \cite{QRS} to the frontal slice $\hat{\mathcal{A}}(:,:,i)$ parallelly with complexity $\mathcal{O}({\rm polylog}(N_1N_2)/{\epsilon_{\rm SVE}})$, where $\epsilon_{\rm SVE}$ is the minimum precision of estimated singular values of $\hat{\mathcal{A}}(:,:,i)$, $i=0,\cdots,N_3-1$. Traditionally, the quantum singular value decomposition of matrices involves exponentiating non-sparse low-rank matrices and output the superposition state of singular values and their associated singular vectors in time $\mathcal{O}({\rm polylog}(N_1, N_2))$ \cite{QSVD}. However, for achieving polylogarithmic complexity, this Hamiltonian simulation method requires that the matrix to be exponentiated is low-rank and it is difficult to be satisfied in general. In our algorithm, we use the QSVE algorithm proposed in  \cite{QRS}, where the matrix is unnecessarily low-rank, sparse or Hermitian and the output is a superposition state of estimated singular values and their associated singular vectors. However, the original QSVE algorithm proposed in  \cite{QRS} has to be carefully modified to become a useful subroutine in our quantum tensor-svd algorithm. In fact, an important result, Theorem \ref{the QSVE for tensor Ahat}, is developed to address this tricky issue.

In Section \ref{complexity of quantum t-svd section}, we show that the proposed quantum t-svd algorithm  for tensors $\mathcal{A}^{N \times N \times N}$ ($N_1=N_2=N_3=N$ for simplification), Algorithm  \ref{alg:Quantum t-svd},  achieves the complexity of $\mathcal{O}(N{\rm polylog}(N)/{\epsilon_{\rm SVE}})$, a time exponentially faster
than its classical counterpart $\mathcal{O}(N^4)$. In Section \ref{quantum t-svd order-p}, we extend the quantum t-svd algorithm to order-$p$ tensors.

In \cite{QRS}, Kerenidis and Prakash designed a quantum algorithm for recommendation systems modeled by an $m \times n$ preference matrix, which makes recommendations by just sampling from an approximation of the preference matrix. Therefore, the running time is only $\mathcal{O}({\rm poly}(k){\rm polylog}(mn))$ if the preference matrix has a good rank-$k$ approximation. To achieve this, they projected a state corresponding to a user's preferences to the approximated row space spanned by singular vectors with singular values greater than the chosen threshold. After measuring this projected state in a computational basis, they got recommended product index for the input user.

In a recommendation system, the task is to predict a user's preferences for a product and then make recommendations. Most recommendation systems do not take context into account. In contrast, context-aware recommendation systems incorporate the contextual situation of a user to personalized recommendation, i.e., a product is recommended to a user varying with different contexts (time, location, etc.). Thus, taking context into account renders a  dynamic recommendation system of three elements user, product and context, which can be neatly modeled by a third-order tensor. We apply our quantum tensor-svd algorithm to these context-aware recommendation systems. Since the product that a user preferred in a certain context is very likely to affect the recommendation for him/her at other contexts, the t-svd factorization technique suits the problem very well because in t-svd the QFT is performed first to bind a user's preferences in different context together. 

t-svd factorization approaches have been shown to have better performance than other tensor decomposition techniques, such as HOSVD, when applied to facial recognition \cite{Hao2013Facial}. It also has good performance in tensor completion \cite{exact_tensor_completion}. However, the computational cost of the t-svd is too high. Compared with the classical t-svd with high complexity, our quantum t-svd is able to both model the context information and reduce the computational complexity. Indeed, our quantum recommendation systems algorithm provides recommendations for a user $i$ by just measuring the output quantum state corresponding to an approximation of the $i$-th frontal slice of the preference tensor. It is designed based on the low-rank tensor reconstruction using t-svd, that is, the full preference tensor can be approximated by the truncated t-svd of the subsample tensor. We also show that  this new quantum recommendation systems algorithm is exponentially faster than its classical counterpart.

The rest of the paper is organized as follows. The classical t-svd algorithm and several related concepts are introduced in Section \ref{section t-svd}; Section \ref{section QSVE} summarizes the quantum singular value estimation algorithm proposed  in  \cite{QRS}. Section \ref{quantum t-svd} provides our main algorithm, quantum t-svd, and its complexity analysis, then extends this algorithm to order-$p$ tensors. In Section \ref{QRS for tensors}, we propose a quantum algorithm for context-aware recommendation systems, whose performance and complexity are analyzed in Sections \ref{Theoretical analysis} and \ref{complexity of QRS} respectively. We prove that for a specified user $i$ to be recommended products, the output state corresponds to an approximation of this user's preference information. Therefore, measuring the output state in the computational basis is a good recommendation for user $i$ with a high probability. In Section \ref{Section: another truncate}, we consider a tensor completion problem and design a quantum algorithm which is similar to Algorithm \ref{alg:QRS for tensors} but truncation is performed in another way.

{\it Notation.} In this paper, script letters are used to denote tensors. Capital nonscript letters are used to represent matrices, and boldface  lower case letters refer to vectors. For a third-order tensor $\mathcal{A} \in \mathbb{R}^{N_1 \times N_2 \times N_3}$, subtensors are formed when a subset of indices is fixed. Specifically, a tube of size $1 \times 1 \times N_3$ can be regarded as a vector and it is defined by fixing all indices but the last one, e.g., $
\mathcal{A}(i,j,:)$. A slice of a tensor $\mathcal{A}$ can be regarded as a matrix defined by fixing one index, e.g., $\mathcal{A}(i,:,:)$, $\mathcal{A}(:,i,:)$, $\mathcal{A}(:,:,i)$ represent the $i$-th horizontal, lateral, frontal slice respectively. We use $A^{(i)}$ to denote the $i$-th frontal slice $\mathcal{A}(:,:,i), i =0, \cdots, N_3-1$. The $i$-th row of the matrix $A$ is denoted by $A_i$. The tensor after the Fourier transform (the FFT for the classical t-svd or the QFT for the quantum t-svd) along the third mode of $\mathcal{A}$ is denoted by $\hat{\mathcal{A}}$ and its $m$-th frontal slice is $\hat{A}^{(m)}$.

\section{Preliminaries}
In this preliminary section, we first review the definition of t-product and the classical (namely, non-quantum) t-svd algorithm proposed by Kilmer et al. \cite{t-svd_three} in 2011. Then in Section \ref{section QSVE}, we briefly review the quantum singular value estimation algorithm (QSVE) \cite{QRS} proposed by Kerenidis and Prakash \cite{QRS} in 2017.

\subsection{The t-svd algorithm based on t-product} \label{section t-svd}
In this subsection, we first review the definition of circulant convolution between vectors, based on which we present the t-product between tensors, finally we present a t-svd algorithm.


\begin{definition}\label{circtensor}
	Given a vector $\boldsymbol{u} \in \mathbb{R}^N$ and a tensor $\mathcal{B} \in \mathbb{R}^{N_1 \times N_2 \times N_3}$ with frontal slices $B^{(\ell)}$, $\ell=0, \cdots, N_3-1$, the matrices ${\rm circ}(\boldsymbol{u})$ and ${\rm circ}(\mathcal{B})$ are defined as \\
\beqnm
{\rm circ}(\boldsymbol{u})
&\triangleq&
   \begin{bmatrix}
		u_0 & u_{N-1} & \cdots & u_1 \\
		u_1 & u_0     & \cdots & u_2 \\
		\vdots & \vdots & \ddots & \vdots \\
		u_{N-1}& u_{N-2} & \cdots & u_0\\
		\end{bmatrix},		\\
{\rm circ}(\mathcal{B})
&\triangleq&
 \begin{bmatrix}
		B^{(0)} & B^{(N_3-1)} & \cdots & B^{(1)} \\
		B^{(1)} & B^{(0)}     & \cdots & B^{(2)} \\
		\vdots & \vdots & \ddots & \vdots \\
		B^{(N_3-1)}& B^{(N_3-2)} & \cdots & B^{(0)}\\
		\end{bmatrix},
\eeqnm
respectively.	
\end{definition}

\begin{definition} \label{def:circ}
	Let $\boldsymbol{u, v} \in \mathbb{R}^N$. The circular convolution between $\boldsymbol{u}$ and $\boldsymbol{v}$ produces a vector $\boldsymbol{x}$ of the same size, defined as $$ \boldsymbol{x} \equiv \boldsymbol{u} \circledast \boldsymbol{v} \triangleq {\rm circ}(\boldsymbol{u})\boldsymbol{v}.$$
\end{definition}

As a circulant matrix can be diagonalized by means of the Fast Fourier transform (FFT), from \eqref{def:circ} we have ${\rm FFT}(\boldsymbol{x}) = {\rm diag}({\rm FFT}(\boldsymbol{u})){\rm FFT}(\boldsymbol{v}),$ where ${\rm diag}(\boldsymbol{u})$ returns a square diagonal matrix with elements of vector $\boldsymbol{u}$ on the main diagonal. ${\rm FFT}(\boldsymbol{u})$ computes the DFT of $\boldsymbol{u},$ i.e., if $\hat{\boldsymbol{u}}={\rm FFT}(\boldsymbol{u}),$ then $\hat{\boldsymbol{u}}_k=\sum_{j=0}^{N-1}e^{-2i\pi kj/N}\boldsymbol{u}_j$, and it reduces the complexity of computing the DFT from $\mathcal{O}(N^2)$ to only $\mathcal{O}(N \log N)$. The next result formalizes the above discussions.

\bthm \cite{convolution} {\bfseries (Cyclic Convolution Theorem)} \label{circular theorem}
Given $\boldsymbol{u}, \boldsymbol{v} \in \mathbb{R}^N$, let $\boldsymbol{x}=\boldsymbol{u} \circledast \boldsymbol{v}$.  We have
\begin{align}
{\rm FFT}(\boldsymbol{x}) = {\rm FFT}(\boldsymbol{u})\odot {\rm FFT}(\boldsymbol{v}),
\end{align}
where $\odot$ is the Hadamard product.
\ethm

If  a tensor $\mathcal{A} \in \mathbb{R}^{N_1 \times N_2 \times N_3}$ is regarded as an $N_1 \times N_2$ matrix of tubes of dimension $N_3$, whose $(i,j)$-th entry (a tube) is $\mathcal{A}(i,j,:)$, then based on the definition of circular convolution between vectors, the t-product between tensors can be defined.

\begin{definition} \cite{t-svd_three} \label{t-product}
	Let $\mathcal{M} \in \mathbb{R}^{N_1 \times N_2 \times N_3}$ and $\mathcal{N} \in \mathbb
	{R}^{N_2 \times N_4 \times N_3}$. The t-product $\mathcal{M}*\mathcal{N}$ is an $N_1 \times N_4 \times N_3$ tensor, denoted by $\mathcal{A}$, whose $(i,j)$-th tube $\mathcal{A}(i,j,:)$ is the sum of the circular convolution between corresponding tubes in the $i$-th horizontal slice of the tensor $\mathcal{M}$ and the $j$-th lateral slice of the tensor $\mathcal{N},$ i.e.,
	\begin{align}
	\mathcal{A}(i,j,:)=\sum_{k=0}^{N_2-1}\mathcal{M}(i,k,:)\circledast \mathcal{N}(k,j,:).
	\end{align}
\end{definition}
\begin{figure}[H]
	\centering
	\includegraphics[width=\linewidth]{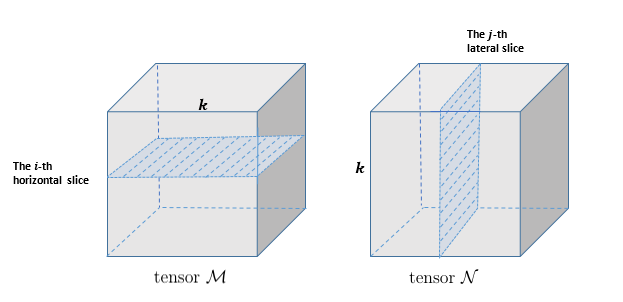}
	\caption{The illustration of the t-product $\mathcal{M}*\mathcal{N}$ in Definition \ref{t-product}}
	\label{the illustration of t-product}
\end{figure}

According to Theorem \ref{circular theorem} and Definition \ref{t-product}, we have
\begin{align}
&{\rm FFT}(\mathcal{A}(i,j,:))
\nonumber
\\
=& \sum_{k=0}^{N_2-1}{\rm FFT}(\mathcal{M}(i,k,:))\odot {\rm FFT}(\mathcal{N}(k,j,:)), \label{tten}
\end{align}
for $i=0, \cdots, N_1-1, \, j=0, \cdots, N_4-1.$ Let $\hat{\mathcal{A}}$ be the tensor, whose $(i,j)$-th tube is ${\rm FFT}(\mathcal{A}(i,j,:))$. Then equation (\ref{tten}) becomes $\hat{\mathcal{A}}(i,j,:)=\sum_{k=0}^{N_2-1}\hat{\mathcal{M}}(i,k,:)\odot \hat{\mathcal{N}}(k,j,:),$ which can also be written in the form $\hat{A}^{(l)}(i,j)=\sum_{k=0}^{N_2-1}\hat{M}^{(l)}(i,k)\hat{N}^{(l)}(k,j)$ for the $\ell$-th frontal slices of these tensors. Therefore, $\hat{A}^{(\ell)}=\hat{M}^{(\ell)}\hat{N}^{(\ell)}$. The following theorem summarizes the idea stated above.

\bthm \cite{t-svd_three}
For any tensor $\mathcal{M} \in \mathbb{R}^{N_1 \times N_2 \times N_3}$ and $\mathcal{N} \in \mathbb{R}^{N_2 \times N_4 \times N_3}$, we have
\begin{align}
\mathcal{A}=\mathcal{M}*\mathcal{N}\quad \Longleftrightarrow \quad  \hat{A}^{(\ell)}=\hat{M}^{(\ell)}\hat{N}^{(l)}
\end{align}
holds for $\ell=0,1,\cdots,N_3-1.$ Moreover, for another tensor $\mathcal{T} \in \mathbb{R}^{N_4 \times N_5 \times N_3}$, we have
\begin{align}
\mathcal{A}=\mathcal{M}*\mathcal{N}*\mathcal{T}\quad \Longleftrightarrow \quad  \hat{A}^{(\ell)}=\hat{M}^{(\ell)}\hat{N}^{(l)}\hat{T}^{(l)},
\end{align} for $\ell=0, \cdots, N_3-1.$
\ethm

Now we can get the tensor decomposition for a tensor $\mathcal{A}$ using the t-product by performing matrix factorization strategies on $\hat{A}^{(\ell)}$. For example, the tensor QR decomposition $\mathcal{A}=\mathcal{Q}*\mathcal{R}$ is defined as performing the matrix QR decomposition on each frontal slice of the tensor $\hat{\mathcal{A}},$ i.e., $\hat{A}^{(\ell)}=\hat{Q}^{(\ell)} \cdot \hat{R}^{(\ell)}$, for $\ell=0, \cdots, N_3-1,$  where $\hat{Q}^{(\ell)}$ is an orthogonal matrix and $\hat{R}^{(\ell)}$ is an upper triangular matrix \cite{t-qr}. If we compute the matrix svd on $\hat{A}^{(\ell)}$, i.e., $\hat{A}^{(\ell)}=\hat{U}^{(\ell)}\hat{S}^{(\ell)}\hat{V}^{(\ell)\dagger}$, the t-svd of tensor $\mathcal{A}$ is obtained; see Algorithm \ref{alg:t-svd}. Before presenting the  t-svd algorithm for third-order tensors, we first introduce some related definitions.

\begin{definition} {\bfseries tensor transpose} \cite{t-svd_three} \label{tensor transpose}\\
	The transpose of a tensor $\mathcal{A} \in \mathbb{R}^{N_1 \times N_2 \times N_3}$, denoted $\mathcal{A}^T$, is obtained by transposing all the frontal slices and then reversing the order of the transposed frontal slices 2 through $N_3$.
\end{definition}

\begin{definition} {\bfseries tensor Frobenius norm} \cite{t-svd_three}\\
	The Frobenius norm of a third-order tensor $\mathcal{A} $ is defined as $||\mathcal{A}||_F=\sqrt{\sum_{i,j,k}|\mathcal{A}(i,j,k)|^2}$.
\end{definition}	

\begin{definition} {\bfseries identity tensor} \cite{t-svd_three}\\
	The identity tensor $\mathcal{I} \in \mathbb{R}^{N_1 \times N_2 \times N_3}$ is a tensor whose first frontal slice $I^{(0)}$ is an $N_1 \times N_1$ identity matrix and all the other frontal slices are zero matrices.
\end{definition}

\begin{definition} {\bfseries orthogonal tensor} \cite{t-svd_three} \label{orthogonal tensor}\\
	A tensor $\mathcal{U} \in \mathbb{R}^{N_1 \times N_2 \times N_3}$ is an orthogonal tensor if it satisfies
	$\mathcal{U}^{T}*\mathcal{U}=\mathcal{U}*\mathcal{U}^{T}=\mathcal{I}$.
\end{definition}

The tensor transpose defined in Definition \ref{tensor transpose} has the same property as the matrix transpose, e.g., $\left(\mathcal{A}*\mathcal{B}\right)^T=\mathcal{B}^T*\mathcal{A}^T$. Similarly, just like orthogonal matrices, the orthogonality defined in Definition \ref{orthogonal tensor} preserves the Frobenius norm of a tensor, i.e., $||\mathcal{Q}*\mathcal{A}||_F=||\mathcal{A}||_F$ if $\mathcal{Q}$ is an orthogonal tensor. Moreover, when the tensor is two-dimensional, Definition \ref{orthogonal tensor} coincides with the definition of orthogonal matrices. Finally, note that frontal slices of an orthogonal tensor are not necessarily orthogonal.

\bthm \cite{t-svd_three} {\bfseries tensor singular value decomposition (t-svd)}  \label{t-svd} \\
For $\mathcal{A} \in \mathbb{R}^{N_1 \times N_2 \times N_3}$, its t-svd is given by $\mathcal{A}=\mathcal{U}*\mathcal{S}*\mathcal{V}^T,$
where $\mathcal{U}\in \mathbb{R}^{N_1 \times N_1 \times N_3},$ $\mathcal{V} \in \mathbb{R}^{N_2 \times N_2 \times N_3}$ are orthogonal tensors, and every frontal slice of $\mathcal{S}\in \mathbb{R}^{N_1 \times N_2 \times N_3}$ is a diagonal matrix.
\ethm

There are several versions of t-svd algorithms. In what follows we present the one proposed in \cite{t-svd_three}.\begin{algorithm}[H]
	\caption{t-svd for third-order tensors \cite{t-svd_three}}
	\label{alg:t-svd}
	\hspace*{\algorithmicindent} \textbf{Input}: $\mathcal{A} \in \mathbb{R}^{N_1 \times N_2 \times N_3}.$\\
	\hspace*{\algorithmicindent} \textbf{Output}: $ \mathcal{U} \in \mathbb{R}^{N_1 \times N_1 \times N_3}, \mathcal{S} \in \mathbb{R}^{N_1 \times N_2 \times N_3}, \mathcal{V} \in \mathbb{R}^{N_2 \times N_2 \times N_3}$
	\begin{algorithmic}
		\STATE $\mathcal{\hat{A}}={\rm fft}(\mathcal{A},[],3);$	
		\FOR{$i=0, \cdots, N_3-1$}
		\STATE $[U, S, V]={\rm svd}(\hat{\mathcal{A}}(:,:,i) );$	
		\STATE $\mathcal{\hat{U}}(:,:,i)=U; \mathcal{\hat{S}}(:,:,i)=S; \mathcal{\hat{V}}(:,:,i)=V;$
		\ENDFOR
		\STATE $\mathcal{U}={\rm ifft}(\mathcal{\hat{U}},[],3); \mathcal{S}={\rm ifft}(\mathcal{\hat{S}},[],3); \mathcal{V}={\rm ifft}(\mathcal{\hat{V}},[],3).$	
	\end{algorithmic}
\end{algorithm}

\begin{remark}
In the t-svd literature, the diagonal elements of the tensor $\mathcal{S}$ are called the singular values of $\mathcal{A}$. Moreover, the $l_2$ norms of the nonzero tubes $\mathcal{S}(i,i,:)$ are in descending order, i.e.,
$||\mathcal{S}(1,1,:)||_2 \geq ||\mathcal{S}(2,2,:)||_2 \geq \cdots \geq ||\mathcal{S}(\min(N_1,N_2),\min(N_1,N_2),:)||_2$. However, it can be noticed that the diagonal elements of $\mathcal{S}$ may  be unordered and even negative due to the inverse FFT.  As a result, when doing tensor truncation in Section  \ref{QRS for tensors} to get quantum recommendation systems, we use $\hat{\mathcal{S}}$ instead of $\mathcal{S}$  as the diagonal elements of the former are non-negative and ordered in descending order.
\end{remark}

Next, we present the definition of the tensor nuclear norm (TNN) which is frequently used as an objective function to be minimized in many optimization algorithms for data completion \cite{exact_tensor_completion}-\cite{Tensor_factorization_for_low-rank}. Since directly minimizing the tensor multi-rank (defined as a vector whose $i$-th entry is the rank of $\hat{A}^{(i)}$) is NP-hard, some works approximate the rank function by its convex surrogate, i.e., TNN \cite{Novel}. It is proved that TNN is the tightest convex relaxation to $\ell_1$ norm of the tensor multi-rank \cite{Novel} and the problem is reduced to a convex one when transformed into minimizing TNN.

\begin{definition} \label{def:tnn}\cite{Novel} {\bfseries Tensor nuclear norm}   \\
	The tensor nuclear norm (TNN) of $\mathcal{A} \in \mathbb{R}^{N_1 \times N_2 \times N_3}$, denoted by $||\mathcal{A}||_{TNN}$, is defined as the sum of the singular values of $\hat{A}^{(\ell)}$, the $\ell$-th frontal slice of $\hat{\mathcal{A}}$, i.e., $||\mathcal{A}||_{TNN}=\sum_{\ell=0}^{N_3-1}||\hat{A}^{(\ell)}||_{*}$, where $||\cdot||_{*}$ refers to the matrix nuclear norm, namely the sum of the singular values.
\end{definition}

An important application of the t-svd algorithm is the optimality of the truncated t-svd for data approximation, which is the theoretical basis of our quantum algorithm for recommendation systems and tensor completion to be developed in Sections \ref{Section QRS for tensors} and \ref{Section: another truncate} respectively. This property is stated in the following Lemma.

\begin{lemma} \cite{t-svd_three,randomized} \label{theoretical basis for QRS}
	Suppose the t-svd of the tensor $\mathcal{A} \in \mathbb{R}^{N_1 \times N_2 \times N_3}$ is $\mathcal{A}=\mathcal{U}*\mathcal{S}*\mathcal{V}^T$. Then we have $$\mathcal{A} = \sum_{i=0}^{\min (N_1,N_2)-1}\mathcal{U}(:,i,:)*\mathcal{S}(i,i,:)*\mathcal{V}(:,i,:)^T,$$ where the matrices $\mathcal{U}(:,i,:)$ and $\mathcal{V}(:,i,:)$ and the vector $\mathcal{S}(i,i,:)$ are regarded as  tensors of order 3. For $1 \leq k < \min (N_1, N_2)$ define $\mathcal{A}_k \triangleq \sum_{i=0}^{k-1}\mathcal{U}(:,i,:)*\mathcal{S}(i,i,:)*\mathcal{V}(:,i,:)^T$. Then
	$$\mathcal{A}_k={\rm arg} \min_{\tilde{\mathcal{A}} \in \mathcal{M}_k}{||\mathcal{A}-\tilde{\mathcal{A}}||_F},$$
	where $\mathcal{M}_k=\{\mathcal{X}*\mathcal{Y}|\mathcal{X} \in \mathbb{R}^{N_1 \times k \times N_3}, \mathcal{Y} \in \mathbb{R}^{k \times N_2 \times N_3}\}$. Therefore, $||\mathcal{A}-\mathcal{A}_k||_F$ is the theoretical minimal error, given by $||\mathcal{A}-\mathcal{A}_k||_F=\sqrt{\sum_{i=k}^{\min (N_1,N_2)-1}||\mathcal{S}(i,i,:)||_2^2}$.
\end{lemma}

\subsection{Quantum singular value estimation}	 \label{section QSVE}

Kerenidis and Prakash \cite{QRS} proposed a quantum algorithm to estimate the singular values of a matrix, named by the quantum singular value estimation (QSVE).  With the introduction of a data structure, see Lemma \ref{data structure} below,  in which the rows of the matrix are stored, the QSVE algorithm can prepare the quantum states corresponding to the rows of the matrix efficiently.
\begin{lemma} \label{data structure} \cite{QRS}
	Consider a matrix $A \in \mathbb{R}^{N_1\times N_2}$ with $\omega$ nonzero entries. Let $A_i$ be its $i$-th row, and $\boldsymbol{s}_A=\frac{1}{||A||_F}\left[||A_0||_2 , ||A_1||_2 , \cdots , ||A_{N_1-1}||_2 \right]^{T}.$ There exists a data structure storing the matrix $A$ in $\mathcal{O}(\omega {\rm log}^2(N_1N_2))$ space such that a quantum algorithm having access to this data structure can perform the mapping $U_{P}: \ket{i}\ket{0}\rightarrow \ket{i}\ket{A_i}$, for $i=0,\cdots, N_1-1$ and $U_{Q}: \ket{0}\ket{j}\rightarrow \ket{\boldsymbol{s}_A}\ket{j}$, for $j=0,\cdots, N_2-1$ in time ${\rm polylog}(N_1N_2)$.	
\end{lemma}

The explicit description of the QSVE is given in \cite{QRS} and the following lemma summarizes the main ideas. Unlike the singular value decomposition technique proposed in Ref. \cite{QPCA, QSVD} that requires the matrix $A$ to be exponentiated be low-rank, in the QSVE algorithm the matrix $A$ is not necessarily sparse or low-rank.
\begin{lemma} \cite{QRS} \label{QSVE}
	Let $A \in \mathbb{R}^{N_1 \times N_2}$ and $\boldsymbol{x} \in \mathbb{R}^{N_2}$ be stored in the data structure  as mentioned in Lemma \ref{data structure}. Let the singular value decomposition of $A$ be $A = \sum_{\ell=0}^{r-1}\sigma_{\ell}\ket{u_{\ell}}\bra{v_{\ell}}$, where $r=\min (N_1, N_2)$. The input state $\ket{x}$ can be represented in the eigenstates of $A$, i.e. $\ket{x}=\sum_{\ell=0}^{N_2-1}\beta_\ell\ket{v_\ell}$. Let $\epsilon > 0$ be the precision parameter. Then there is a quantum algorithm, denoted as $U_{{\rm SVE}}$, that runs in $\mathcal{O}({\rm polylog}(N_1N_2)/{\epsilon})$ and achieves $$U_{{\rm SVE}}\left( \ket{x}\ket{0}\right)=\sum_{\ell=0}^{N_2-1}\beta_\ell\ket{v_\ell}\ket{\overline{\sigma}_{\ell}},$$ where $\overline{\sigma}_{\ell}$ is the estimated value of $\sigma_{\ell}$ satisfying
	$|\overline{\sigma}_{\ell}-\sigma_{\ell}| \leq \epsilon||A||_F$ for all $\ell$ with probability at least $1-1/{\rm poly}(N_2)$.
\end{lemma}


\begin{remark} \label{modified QSVE remarks}	
In the QSVE algorithm on the matrix $A$ stated in Lemma \ref{QSVE},  we can also choose the input state as
	 $\ket{A}=\frac{1}{||A||_F}\sum_{\ell=0}^{r-1}\sigma_{\ell}\ket{u_\ell} \ket{v_\ell}$, corresponding to the vectorized form of the normalized matrix \\
	$\frac{A}{||A||_F}=\frac{1}{||A||_F}\sum_{i,j}a_{ij}\ket{i}\bra{j}$ represented in the svd form. This representation of the input state is adopted in Section \ref{quantum t-svd}. Note that we can express the state $\ket{A}$ in the above form even if we do not know the singular pairs of $A$. According to Lemma \ref{QSVE}, we can obtain $\overline{\sigma}_{\ell}$, an estimation of $\sigma_\ell$, stored in the third register superposed with the singular pair $\{\ket{u_\ell}, \ket{v_\ell}\}$ after performing $U_{{\rm SVE}}$, i.e., the output state is $\frac{1}{||A||_F}\sum_{\ell=0}^{r-1}\sigma_{\ell}\ket{u_\ell} \ket{v_\ell}\ket{\overline{\sigma}_\ell}$, where $|\overline{\sigma}_{\ell} - \sigma_{\ell}| \leq \epsilon ||A||_F$ for all $\ell=0, \cdots,r-1$.
\end{remark}

\section{Quantum t-svd algorithms} \label{quantum t-svd}

In this section, we present our quantum t-svd algorithm for third-order tensors. We also show that the running time of this algorithm is exponentially faster than its classical counterpart, provided that every frontal slice of the tensor is stored in the  data structure as introduced in Lemma \ref{data structure} and the tensor as a quantum state can be efficiently prepared. We first present the algorithm (Algorithm \ref{alg:Quantum t-svd}) in Section \ref{quantum t-svd section},  then we analyze its computational complexity in Section \ref{complexity of quantum t-svd section}. Finally, we extend it to order-$p$ tensors in Section \ref{quantum t-svd order-p}.

For a third-order tensor $\mathcal{A} \in \mathbb{R}^{N_1\times N_2 \times N_3},$ we assume that every frontal slice of $\mathcal{A}$ is stored in a tree structure introduced in Lemma \ref{data structure} such that the algorithm having quantum access to this data structure can return the desired quantum state.

\begin{assumption} \label{model assumption}
	Let tensor $\mathcal{A} \in \mathbb{R}^{N_1\times N_2 \times N_3}$, where $N_i=2^{n_i}$ with $n_i$ being the number of qubits on the $i$-th mode, $i=1, 2, 3$. Assume that we have an efficient quantum algorithm (e.g. QRAM) to achieve the quantum state preparation
	\begin{align} \label{state tensor A}
	\ket{\mathcal{A}}= \frac{1}{||\mathcal{A}||_F}\sum_{i=0}^{N_1-1}\sum_{j=0}^{N_2-1}\sum_{k=0}^{N_3-1}\mathcal{A}(i,j,k)\ket{i}\ket{j}\ket{k}
	\end{align} efficiently. That is, we can encode $\mathcal{A}(i,j,k)$ as the amplitude of a three-partite system. Without loss of generality, we assume that $||\mathcal{A}||_F=1$.
\end{assumption}

\subsection{Quantum t-svd for third-order tensors} \label{quantum t-svd section}

In this section, we first present our quantum t-svd algorithm, Algorithm \ref{alg:Quantum t-svd}, for third-order tensors, then explain each step in detail.

\begin{algorithm}[H]
	\caption{Quantum t-svd for third-order tensors }
	\label{alg:Quantum t-svd}
	\hspace*{\algorithmicindent} \textbf{Input:} tensor $\mathcal{A} \in \mathbb{R}^{N_1\times N_2 \times N_3}$ prepared in a quantum state $\ket{\mathcal{A}}$ in (\ref{state tensor A}), precision $\epsilon_{\rm SVE}^{(m)}$, $m=0, \cdots, N_3-1$, $r=\min\{N_1, N_2\}$. \\
	\hspace*{\algorithmicindent} \textbf{Output:} the state $\ket{\phi}.$
	\begin{algorithmic}[1]
		\STATE Perform the QFT on the third register of the quantum state $\ket{\mathcal{A}}$, to obtain the state $\ket{\mathcal{\hat{A}}}$.
		\STATE Perform the controlled-$U_{{\rm SVE}}$ on the state $\ket{\mathcal{\hat{A}}}$ to get the state
\beq\label{eq:psi_jan13}
\ket{\psi}=\sum_{m=0}^{N_3-1}\left( \sum_{\ell=0}^{r-1}\hat{\sigma}_\ell^{(m)}\ket{\hat{u}_\ell^{(m)}}^{c}\ket{\hat{v}_\ell^{(m)}}^{d}\ket{\overline{\hat{\sigma}}_\ell^{(m)}}^{a} \right)\ket{m}^{e}.
\eeq
		\STATE Perform the inverse QFT on the last register of $\ket{\psi}$ and output the state
\beq\label{eq:phi_jan13}		
		\ket{\phi}=\frac{1}{\sqrt{N_3}}\sum_{t,m=0}^{N_3-1} \sum_{\ell=0}^{r-1}\hat{\sigma}_i^{(m)}\omega^{-tm}\ket{\hat{u}_\ell^{(m)}}^{c}\ket{\hat{v}_\ell^{(m)}}^{d}\ket{\overline{\hat{\sigma}}_\ell^{(m)}}^{a} \ket{t}^{e}.
\eeq
	\end{algorithmic}
\end{algorithm}	

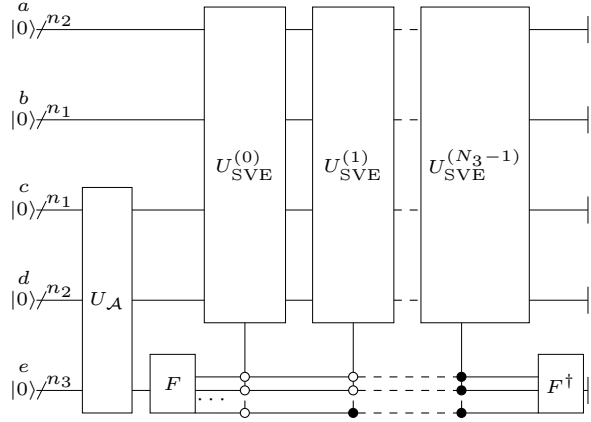
\begin{figure}[H]
\begin{center}
	\begin{tikzpicture}[scale=1.2, transform shape]
	\tikzset{font=\small}
	\tikzstyle{basicshadow}=[blur shadow={shadow blur steps=8, shadow xshift=0.7pt, shadow yshift=-0.7pt, shadow scale=1.02}]\tikzstyle{basic}=[draw,fill=white,basicshadow]
	\tikzstyle{operator}=[text width=3em, text centered, minimum height=3em]
	\tikzstyle{phase}=[fill=white,shape=circle,minimum size=0.1cm,inner sep=0pt,outer sep=0pt,draw=black]
	\tikzstyle{phase1}=[fill=black,shape=circle,minimum size=0.1cm,inner sep=0pt,outer sep=0pt,draw=black]
	\tikzstyle{none}=[inner sep=0pt,outer sep=-.5pt, minimum height=0.5cm+1pt]
	\tikzstyle{measure}=[operator,inner sep=0pt,minimum height=0.5cm, minimum width=0.75cm]
	\tikzstyle{xstyle}=[circle,basic,minimum height=0.35cm,minimum width=0.35cm,inner sep=-1pt,very thin]
	\tikzset{
		shadowed/.style={preaction={transform canvas={shift={(0.5pt,-0.5pt)}}, draw=gray, opacity=0.3}},
	}
	\tikzstyle{swapstyle}=[inner sep=-2pt, outer sep=-1pt, minimum width=0pt]
	\tikzstyle{edgestyle}=[very thin]
	
	\node[none] (line0_gate0) at (0.1,-0) {\tiny $\Ket{0}$};

	\node[none] (line0_gate45) at (0.5,0.1) {\tiny $n_2$};
	\node[none] (line1_gate45) at (0.5,-0.9) {\tiny $n_1$};
	\node[none] (line2_gate45) at (0.5,-1.9) {\tiny $n_1$};
	\node[none] (line3_gate45) at (0.5,-2.9) {\tiny $n_2$};
	\node[none] (line4_gate45) at (0.5,-3.9) {\tiny $n_3$};
	
	\node[none] (line0_gate145) at (0.1,0.25) {\tiny $a$};
	\node[none] (line1_gate145) at (0.1,-0.75) {\tiny $b$};
	\node[none] (line2_gate145) at (0.1,-1.75) {\tiny $c$};
	\node[none] (line3_gate145) at (0.1,-2.75) {\tiny $d$};
	\node[none] (line4_gate145) at (0.1,-3.75) {\tiny $e$};
	
	\node[none] (line1_gate0) at (0.1,-1) {\tiny $\Ket{0}$};
	\node[none] (line2_gate0) at (0.1,-2) {\tiny $\Ket{0}$};
	\node[none] (line3_gate0) at (0.1,-3) {\tiny $\Ket{0}$};
	\node[none] (line4_gate0) at (0.1,-4) {\tiny $\Ket{0}$};
	\node[none] (line2_gate1) at (0.75,-2) {};
	\node[none,minimum height=0.5cm,outer sep=0] (line2_gate2) at (0.9,-2) {};
	\node[none] (line2_gate3) at (1.3,-2) {};
	\node[none] (line3_gate1) at (0.75,-3) {};
	\node[none,minimum height=0.5cm,outer sep=0] (line3_gate2) at (0.9,-3) {};
	\node[none] (line3_gate3) at (1.3,-3) {};
	\node[none] (line4_gate1) at (0.75,-4) {};
	\node[none,minimum height=0.5cm,outer sep=0] (line4_gate2) at (0.9,-4) {};
	\node[none] (line4_gate3) at (1.3,-4) {};
	\draw[operator,edgestyle,outer sep=0.5cm] ([yshift=0.25cm]line2_gate1) rectangle ([yshift=-0.25cm]line4_gate3) node[pos=.5] {\tiny $ U_{\tiny \mathcal{A}}$};
	\draw (line2_gate0) edge[edgestyle] (line2_gate1);
	\draw (line3_gate0) edge[edgestyle] (line3_gate1);
	\draw (line4_gate0) edge[edgestyle] (line4_gate1);
	\node[none] (line4_gate4) at (1.5,-4) {};
	\node[none,minimum height=0.5cm,outer sep=0] (line4_gate5) at (1.75,-4) {};
	\node[none] (line4_gate6) at (2.0,-3.85) {};
	\draw[operator,edgestyle,outer sep=0.5cm] ([yshift=0.4cm]line4_gate4) rectangle ([yshift=-0.4cm]line4_gate6) node[pos=.5] {\tiny $F$};
	
	\draw (line4_gate3) edge[edgestyle] (line4_gate4);
	\node[none] (line0_gate4) at (2.1000000000000003,-0) {};
	\node[none,minimum height=0.5cm,outer sep=0] (line0_gate5) at (2.5500000000000003,-0) {};
	\node[none] (line0_gate6) at (3.000000000000003,-0) {};
	\node[none] (line1_gate1) at (2.1000000000000003,-1) {};
	\node[none,minimum height=0.5cm,outer sep=0] (line1_gate2) at (2.5500000000000003,-1) {};
	\node[none] (line1_gate3) at (3.000000000000003,-1) {};
	\node[none] (line2_gate4) at (2.1000000000000003,-2) {};
	\node[none,minimum height=0.5cm,outer sep=0] (line2_gate5) at (2.5500000000000003,-2) {};
	\node[none] (line2_gate6) at (3.000000000000003,-2) {};
	\node[none] (line3_gate4) at (2.1000000000000003,-3) {};
	\node[none,minimum height=0.5cm,outer sep=0] (line3_gate5) at (2.5500000000000003,-3) {};
	\node[none] (line3_gate6) at (3.000000000000003,-3) {};
	\draw[operator,edgestyle,outer sep=0.5cm] ([yshift=0.25cm]line0_gate4) rectangle ([yshift=-0.25cm]line3_gate6) node[pos=.5] {\tiny $U_{\rm SVE}^{(0)}$};
	\node[phase] (line4_gate7) at (2.5500000000000003,-3.85) {};
	\draw (line4_gate7) edge[edgestyle] (line3_gate5);
	\draw (line0_gate0) edge[edgestyle] (line0_gate4);
	\draw (line1_gate0) edge[edgestyle] (line1_gate1);
	\draw (line2_gate3) edge[edgestyle] (line2_gate4);
	\draw (line3_gate3) edge[edgestyle] (line3_gate4);
	\draw (line4_gate6) edge[edgestyle] (line4_gate7);
	\node[none] (line0_gate7) at (3.3000000000000003,-0) {};
	\node[none,minimum height=0.5cm,outer sep=0] (line0_gate8) at (3.7500000000000003,-0) {};
	\node[none] (line0_gate9) at (4.2000000000000003,-0) {};
	\node[none] (line1_gate4) at (3.3000000000000003,-1) {};
	\node[none,minimum height=0.5cm,outer sep=0] (line1_gate5) at (3.7500000000000003,-1) {};
	\node[none] (line1_gate6) at (4.2000000000000003,-1) {};
	\node[none] (line2_gate7) at (3.3000000000000003,-2) {};
	\node[none,minimum height=0.5cm,outer sep=0] (line2_gate8) at (3.7500000000000003,-2) {};
	\node[none] (line2_gate9) at (4.2000000000000003,-2) {};
	\node[none] (line3_gate7) at (3.3000000000000003,-3) {};
	\node[none,minimum height=0.5cm,outer sep=0] (line3_gate8) at (3.7500000000000003,-3) {};
	\node[none] (line3_gate9) at (4.2000000000000003,-3) {};
	\draw[operator,edgestyle,outer sep=0.5cm] ([yshift=0.25cm]line0_gate7) rectangle ([yshift=-0.25cm]line3_gate9) node[pos=.5] {\tiny $U_{\rm SVE}^{(1)}$};
	\node[phase] (line4_gate8) at (3.7500000000000003,-3.85) {};
	\draw (line4_gate8) edge[edgestyle] (line3_gate8);
	\draw (line0_gate6) edge[edgestyle] (line0_gate7);
	\draw (line1_gate3) edge[edgestyle] (line1_gate4);
	\draw (line2_gate6) edge[edgestyle] (line2_gate7);
	\draw (line3_gate6) edge[edgestyle] (line3_gate7);
	\draw (line4_gate7) edge[edgestyle] (line4_gate8);
	\node[none] (line0_gate10) at (4.5,-0) {};
	\node[none,minimum height=0.5cm,outer sep=0] (line0_gate11) at (4.95,-0) {};
	\node[none] (line0_gate12) at (5.7,-0) {};
	\node[none] (line1_gate7) at (4.5,-1) {};
	\node[none,minimum height=0.5cm,outer sep=0] (line1_gate8) at (4.95,-1) {};
	\node[none] (line1_gate9) at (5.7,-1) {};
	\node[none] (line2_gate10) at (4.5,-2) {};
	\node[none,minimum height=0.5cm,outer sep=0] (line2_gate11) at (4.95,-2) {};
	\node[none] (line2_gate12) at (5.7,-2) {};
	\node[none] (line3_gate10) at (4.5,-3) {};
	\node[none,minimum height=0.5cm,outer sep=0] (line3_gate11) at (4.95,-3) {};
	\node[none] (line3_gate12) at (5.7,-3) {};
	\draw[operator,edgestyle,outer sep=0.5cm] ([yshift=0.25cm]line0_gate10) rectangle ([yshift=-0.25cm]line3_gate12) node[pos=.5] {\tiny $U_{\rm SVE}^{(N_3-1)}$};
	\node[phase1] (line4_gate9) at (4.95,-3.85) {};
	\draw (line4_gate9) edge[edgestyle] (line3_gate11);
	\draw (line0_gate9) edge[dashed] (line0_gate10);
	\draw (line1_gate6) edge[dashed] (line1_gate7);
	\draw (line2_gate9) edge[dashed] (line2_gate10);
	\draw (line3_gate9) edge[dashed] (line3_gate10);
	\draw (line4_gate8) edge[dashed] (line4_gate9);
	\node[none] (line0_gate13) at (6.35,-0) {};
	\draw ([yshift=0.15cm]line0_gate13.center) edge [edgestyle] ([yshift=-0.15cm]line0_gate13.center);
	\draw (line0_gate12) edge[edgestyle] (line0_gate13);
	\node[none] (line1_gate10) at (6.35,-1) {};
	\draw ([yshift=0.15cm]line1_gate10.center) edge [edgestyle] ([yshift=-0.15cm]line1_gate10.center);
	\draw (line1_gate9) edge[edgestyle] (line1_gate10);
	\node[none] (line2_gate13) at (6.35,-2) {};
	\draw ([yshift=0.15cm]line2_gate13.center) edge [edgestyle] ([yshift=-0.15cm]line2_gate13.center);
	\draw (line2_gate12) edge[edgestyle] (line2_gate13);
	\node[none] (line3_gate13) at (6.35,-3) {};
	\draw ([yshift=0.15cm]line3_gate13.center) edge [edgestyle] ([yshift=-0.15cm]line3_gate13.center);
	\draw (line3_gate12) edge[edgestyle] (line3_gate13);
	\node[none] (line4_gate10) at (5.8,-3.85) {};
	\node[none,minimum height=0.5cm,outer sep=0] (line4_gate11) at (6.05,-4) {};
	\node[none] (line4_gate12) at (6.3,-4) {};
	\draw[operator,edgestyle,outer sep=0.5cm] ([yshift=0.25cm]line4_gate10) rectangle ([yshift=-0.25cm]line4_gate12) node[pos=.5] {\tiny $F^{\dagger}$};
	\draw (line4_gate9) edge[edgestyle] (line4_gate10);
	\node[none] (line4_gate13) at (6.35,-4) {};
	\draw ([yshift=0.15cm]line4_gate13.center) edge [edgestyle] ([yshift=-0.15cm]line4_gate13.center);
	\draw (line4_gate12) edge[edgestyle] (line4_gate13);
	
	\node[none] (line4_gate14) at (2,-4) {};
	\node[none] (line4_gate15) at (2.55,-4) {};
	\node[none] (line4_gate16) at (3.75,-4) {};
	\node[none] (line4_gate17) at (4.95,-4) {};
	\node[none] (line4_gate18) at (5.8,-4) {};
	\node[phase] (line4_gate15) at (2.55,-4) {};
	\node[phase] (line4_gate16) at (3.75,-4) {};
	\node[phase1] (line4_gate17) at (4.95,-4) {};
	\draw (line4_gate14) edge[edgestyle] (line4_gate15);
	\draw (line4_gate15) edge[edgestyle] (line4_gate16);
	\draw (line4_gate16) edge[dashed] (line4_gate17);
	\draw (line4_gate17) edge[edgestyle] (line4_gate18);
	\draw (line4_gate15) edge[edgestyle] (line4_gate7);
	\draw (line4_gate16) edge[edgestyle] (line4_gate8);
	\draw (line4_gate17) edge[edgestyle] (line4_gate9);
	
	\node[none] (line0_gate40) at (0.25,-0.1){};
	\node[none] (line0_gate41) at (0.35,0.1){};
	\draw (line0_gate40) edge[edgestyle] (line0_gate41);
	
	\node[none] (line1_gate40) at (0.25,-1.1){};
	\node[none] (line1_gate41) at (0.35,-0.9){};
	\draw (line1_gate40) edge[edgestyle] (line1_gate41);
	
	\node[none] (line2_gate40) at (0.25,-2.1){};
	\node[none] (line2_gate41) at (0.35,-1.9){};
	\draw (line2_gate40) edge[edgestyle] (line2_gate41);
	
	\node[none] (line3_gate40) at (0.25,-3.1){};
	\node[none] (line3_gate41) at (0.35,-2.9){};
	\draw (line3_gate40) edge[edgestyle] (line3_gate41);
	
	\node[none] (line4_gate40) at (0.25,-4.1){};
	\node[none] (line4_gate41) at (0.35,-3.9){};
	\draw (line4_gate40) edge[edgestyle] (line4_gate41);
	
	\node[none] (line4_gate19) at (2,-4.25) {};
	\node[none] (line4_gate20) at (2.55,-4.25) {};
	\node[none] (line4_gate21) at (3.75,-4.25) {};
	\node[none] (line4_gate22) at (4.95,-4.25) {};
	\node[none] (line4_gate23) at (5.8,-4.25) {};
	\node[phase] (line4_gate20) at (2.55,-4.25) {};
	\node[phase1] (line4_gate21) at (3.75,-4.25) {};
	\node[phase1] (line4_gate22) at (4.95,-4.25) {};
	\draw (line4_gate19) edge[edgestyle] (line4_gate20);
	\draw (line4_gate20) edge[edgestyle] (line4_gate21);
	\draw (line4_gate21) edge[dashed] (line4_gate22);
	\draw (line4_gate22) edge[edgestyle] (line4_gate23);
	\draw (line4_gate20) edge[dashed] (line4_gate15);
	\draw (line4_gate21) edge[dashed] (line4_gate16);
	\draw (line4_gate22) edge[dashed] (line4_gate17);
	
	\node[none] (line4_gate60) at (2.2,-4.1) {};
	\node[none] (line4_gate61) at (2.4,-4.1) {};
	\path[] (line4_gate60) node { \tiny $\cdots$ } (line4_gate61);
	
	\end{tikzpicture}
	\caption{Circuit of Algorithm \ref{alg:Quantum t-svd}. $U_{\mathcal{A}}$ is the unitary operator for preparing the state $\ket{\mathcal{A}}$. The QFT is denoted by $F.$ The blocks $U_{\rm SVE}^{(m)}$ are further illustrated in FIG. \ref{Fig.3}.} \label{Fig.2}
	\end{center}
\end{figure}

\begin{figure}[H]
\begin{center}
	\begin{tikzpicture}[scale=1.1, transform shape]
	\tikzstyle{basicshadow}=[blur shadow={shadow blur steps=8, shadow xshift=0.7pt, shadow yshift=-0.7pt, shadow scale=1.02}]\tikzstyle{basic}=[draw,fill=white,basicshadow]
	\tikzstyle{operator}=[basic,minimum size=1.5em]
	\tikzstyle{phase}=[fill=black,shape=circle,minimum size=0.1cm,inner sep=0pt,outer sep=0pt,draw=black]
	\tikzstyle{none}=[inner sep=0pt,outer sep=-.5pt,minimum height=0.5cm+1pt]
	\tikzstyle{measure}=[operator,inner sep=0pt,minimum height=0.5cm, minimum width=0.75cm]
	\tikzstyle{xstyle}=[circle,basic,minimum height=0.35cm,minimum width=0.35cm,inner sep=-1pt,very thin]
	\tikzset{
		shadowed/.style={preaction={transform canvas={shift={(0.5pt,-0.5pt)}}, draw=gray, opacity=0.4}},
	}
	\tikzstyle{swapstyle}=[inner sep=-1pt, outer sep=-1pt, minimum width=0pt]
	\tikzstyle{edgestyle}=[very thin]
	
	\node[none] (line0_gate0) at (0.1,-0) {\tiny $\Ket{0}$};
	\node[none] (line0_gate1) at (0.6,-0) {};
	\node[none,minimum height=0.5cm,outer sep=0] (line0_gate2) at (0.85,-0) {};
	\node[none] (line0_gate3) at (1,0.07) {};
	\draw[operator,edgestyle,outer sep=0.5cm] ([yshift=0.25cm]line0_gate1) rectangle ([yshift=-0.4cm]line0_gate3) node[pos=.5] {\tiny H};
	\draw (line0_gate0) edge[edgestyle] (line0_gate1);
	\node[none] (line1_gate0) at (0.1,-1) {\tiny $\Ket{0}$};
	\node[none] (line2_gate0) at (0.1,-2) {\tiny $\Ket{0}$};
	\node[none] (line3_gate0) at (0.1,-3) {\tiny $\Ket{0}$};
	\node[swapstyle] (line2_gate1) at (0.75,-2) {};
	\draw[swapstyle,edgestyle,shadowed] ([xshift=-0.1cm,yshift=-0.1cm]line2_gate1.center)--([xshift=0.1cm,yshift=0.1cm]line2_gate1.center);
	\draw[swapstyle,edgestyle,shadowed] ([xshift=-0.1cm,yshift=0.1cm]line2_gate1.center)--([xshift=0.1cm,yshift=-0.1cm]line2_gate1.center);
	\node[swapstyle] (line3_gate1) at (0.75,-3) {};
	\draw[swapstyle,edgestyle,shadowed] ([xshift=-0.1cm,yshift=-0.1cm]line3_gate1.center)--([xshift=0.1cm,yshift=0.1cm]line3_gate1.center);
	\draw[swapstyle,edgestyle,shadowed] ([xshift=-0.1cm,yshift=0.1cm]line3_gate1.center)--([xshift=0.1cm,yshift=-0.1cm]line3_gate1.center);
	\draw (line2_gate1) edge[edgestyle] (line3_gate1);
	\draw (line2_gate0) edge[edgestyle] (line2_gate1);
	\draw (line3_gate0) edge[edgestyle] (line3_gate1);
	\node[none] (line1_gate1) at (1.05,-1) {};
	\node[none,minimum height=0.5cm,outer sep=0] (line1_gate2) at (1.4,-1) {};
	\node[none] (line1_gate3) at (1.8,-1) {};
	\node[none] (line2_gate2) at (1.05,-2) {};
	\node[none,minimum height=0.5cm,outer sep=0] (line2_gate3) at (1.4,-2) {};
	\node[none] (line2_gate4) at (1.8,-2) {};
	\draw[operator,edgestyle,outer sep=0.5cm] ([yshift=0.25cm]line1_gate1) rectangle ([yshift=-0.25cm]line2_gate4) node[pos=.5] {\tiny $\hat{U}_Q^{(m)}$};
	\draw (line1_gate0) edge[edgestyle] (line1_gate1);
	\draw (line2_gate1) edge[edgestyle] (line2_gate2);
	\node[none] (line1_gate4) at (1.900000000000002,-1) {};
	\node[none,minimum height=0.5cm,outer sep=0] (line1_gate5) at (2.2,-1) {};
	\node[none] (line1_gate6) at (2.45,-1) {};
	\node[none] (line2_gate5) at (1.900000000000002,-2) {};
	\node[none,minimum height=0.5cm,outer sep=0] (line2_gate6) at (2.2,-2) {};
	\node[none] (line2_gate7) at (2.45,-2) {};
	\draw[operator,edgestyle,outer sep=0.5cm] ([yshift=0.25cm]line1_gate4) rectangle ([yshift=-0.25cm]line2_gate7) node[pos=.5] {\tiny $W_m^{2^0}$};
	\node[phase] (line0_gate4) at (2.2,-0.32) {};
	\draw (line0_gate4) edge[edgestyle] (line1_gate5);
	\draw (line1_gate3) edge[edgestyle] (line1_gate4);
	\draw (line2_gate4) edge[edgestyle] (line2_gate5);

	\node[none] (line0_gate30) at (0.25,-0.1){};
	\node[none] (line0_gate31) at (0.35,0.1){};
	\draw (line0_gate30) edge[edgestyle] (line0_gate31);
	
	\node[none] (line1_gate30) at (0.25,-1.1){};
	\node[none] (line1_gate31) at (0.35,-0.9){};
	\draw (line1_gate30) edge[edgestyle] (line1_gate31);
	
	\node[none] (line2_gate30) at (0.25,-2.1){};
	\node[none] (line2_gate31) at (0.35,-1.9){};
	\draw (line2_gate30) edge[edgestyle] (line2_gate31);
	
	\node[none] (line3_gate30) at (0.25,-3.1){};
	\node[none] (line3_gate31) at (0.35,-2.9){};
	\draw (line3_gate30) edge[edgestyle] (line3_gate31);
	
	\node[none] (line0_gate23) at (1,0.2) {};
	\node[phase] (line0_gate6) at (4.0,0.2) {};
	
	\node[none] (line0_gate27) at (4.75,0.2) {};
	\draw (line0_gate6) edge[edgestyle] (line0_gate27);
	\node[phase] (line0_gate5) at (2.95,0.07) {};
	\node[none] (line0_gate7) at (4.750000000000001,0.07) {};
	\draw (line0_gate3) edge[edgestyle] (line0_gate5);

	\node[none] (line0_gate18) at (1,-0.32) {};
	\node[none] (line0_gate37) at (4.75,-0.32) {};
	
	\draw (line0_gate4) edge[edgestyle] (line0_gate18);
	
	\node[none] (line1_gate7) at (2.6,-1) {};
	\node[none,minimum height=0.5cm,outer sep=0] (line1_gate8) at (2.95,-1) {};
	\node[none] (line1_gate9) at (3.2,-1) {};
	\node[none] (line2_gate8) at (2.6,-2) {};
	\node[none,minimum height=0.5cm,outer sep=0] (line2_gate9) at (2.95,-2) {};
	\node[none] (line2_gate10) at (3.2,-2) {};
	\draw[operator,edgestyle,outer sep=0.5cm] ([yshift=0.25cm]line1_gate7) rectangle ([yshift=-0.25cm]line2_gate10) node[pos=.5] {\tiny $W_m^{2^1}$};
	\draw (line1_gate6) edge[edgestyle] (line1_gate7);
	\draw (line2_gate7) edge[edgestyle] (line2_gate8);
		
	\node[none] (line1_gate10) at (3.6500000000000004,-1) {};
	\node[none,minimum height=0.5cm,outer sep=0] (line1_gate11) at (4.0,-1) {};
	\node[none] (line1_gate12) at (4.45,-1) {};
	\node[none] (line2_gate11) at (3.6500000000000004,-2) {};
	\node[none,minimum height=0.5cm,outer sep=0] (line2_gate12) at (4.0,-2) {};
	\node[none] (line2_gate13) at (4.45,-2) {};
	\draw[operator,edgestyle,outer sep=0.5cm] ([yshift=0.25cm]line1_gate10) rectangle ([yshift=-0.25cm]line2_gate13) node[pos=.5] {\tiny $W_m^{2^d}$};
	\draw (line0_gate6) edge[edgestyle] (line1_gate11);

	\draw (line1_gate9) edge[dashed] (line1_gate10);
	\draw (line2_gate10) edge[dashed] (line2_gate11);
	\draw (line0_gate5) edge[edgestyle] (line1_gate8);

	\node[none] (line0_gate45) at (0.45,0.1) {\tiny $n_2$};
	\node[none] (line1_gate45) at (0.5,-0.9) {\tiny $n_1$};
	\node[none] (line2_gate45) at (0.5,-1.9) {\tiny $n_1$};
	\node[none] (line3_gate45) at (0.5,-2.9) {\tiny $n_2$};
	
	\node[none] (line0_gate145) at (0.1,0.25) {\tiny $a$};
	\node[none] (line1_gate145) at (0.1,-0.75) {\tiny $b$};
	\node[none] (line2_gate145) at (0.1,-1.75) {\tiny $c$};
	\node[none] (line3_gate145) at (0.1,-2.75) {\tiny $d$};

	\node[none,minimum height=0.5cm,outer sep=0] (line0_gate8) at (5.000000000000001,-0) {};
	\node[none] (line0_gate9) at (5.250000000000001,-0) {};
	\draw[operator,edgestyle,outer sep=0.5cm] ([yshift=0.2cm]line0_gate7) rectangle ([yshift=-0.35cm]line0_gate9) node[pos=.5] {\tiny $F^{\dagger}$};
	
	\node[none] (line0_gate10) at (5.550000000000001,-0) {};
	\node[none,minimum height=0.5cm,outer sep=0] (line0_gate11) at (5.900000000000001,-0) {};
	\node[none] (line0_gate12) at (6.250000000000001,-0) {};
	\draw[operator,edgestyle,outer sep=0.5cm] ([yshift=0.25cm]line0_gate10) rectangle ([yshift=-0.4cm]line0_gate12) node[pos=.5] {\tiny $U_{f_m}$};
	\draw (line0_gate9) edge[edgestyle] (line0_gate10);
	\node[none] (line0_gate13) at (7.550000000000001,-0) {};
	\draw ([yshift=0.15cm]line0_gate13.center) edge [edgestyle] ([yshift=-0.15cm]line0_gate13.center);
	\draw (line0_gate12) edge[edgestyle] (line0_gate13);
	\node[none] (line1_gate13) at (6.3,-1) {};
	\node[none,minimum height=0.5cm,outer sep=0] (line1_gate14) at (7.05,-1) {};
	\node[none] (line1_gate15) at (7.55,-1) {};
	\node[none] (line2_gate14) at (6.3,-2) {};
	\node[none,minimum height=0.5cm,outer sep=0] (line2_gate15) at (6.7,-2) {};
	\node[none] (line2_gate16) at (7.05,-2) {};
	\draw[operator,edgestyle,outer sep=0.5cm] ([yshift=0.25cm]line1_gate13) rectangle ([yshift=-0.25cm]line2_gate16) node[pos=.5] {\tiny $\hat{U}_Q^{(m)\dagger}$};
	\draw (line1_gate12) edge[edgestyle] (line1_gate13);
	\draw (line2_gate13) edge[edgestyle] (line2_gate14);
	
	\node[none] (line1_gate16) at (7.55,-1) {};
	\draw (line1_gate14) edge[edgestyle] (line1_gate16);
	\draw ([yshift=0.15cm]line1_gate16.center) edge [edgestyle] ([yshift=-0.15cm]line1_gate16.center);
	\node[swapstyle] (line2_gate17) at (7.25,-2) {};
	\draw[swapstyle,edgestyle,shadowed] ([xshift=-0.1cm,yshift=-0.1cm]line2_gate17.center)--([xshift=0.1cm,yshift=0.1cm]line2_gate17.center);
	\draw[swapstyle,edgestyle,shadowed] ([xshift=-0.1cm,yshift=0.1cm]line2_gate17.center)--([xshift=0.1cm,yshift=-0.1cm]line2_gate17.center);
	\node[swapstyle] (line3_gate2) at (7.25,-3) {};
	\draw[swapstyle,edgestyle,shadowed] ([xshift=-0.1cm,yshift=-0.1cm]line3_gate2.center)--([xshift=0.1cm,yshift=0.1cm]line3_gate2.center);
	\draw[swapstyle,edgestyle,shadowed] ([xshift=-0.1cm,yshift=0.1cm]line3_gate2.center)--([xshift=0.1cm,yshift=-0.1cm]line3_gate2.center);
	\draw (line2_gate17) edge[edgestyle] (line3_gate2);
	\draw (line2_gate16) edge[edgestyle] (line2_gate17);
	\draw (line3_gate1) edge[edgestyle] (line3_gate2);
	\node[none] (line2_gate18) at (7.55,-2) {};
	\draw ([yshift=0.15cm]line2_gate18.center) edge [edgestyle] ([yshift=-0.15cm]line2_gate18.center);
	\draw (line2_gate18) edge[edgestyle] (line2_gate17);
	\node[none] (line3_gate3) at (7.55,-3) {};
	\draw ([yshift=0.15cm]line3_gate3.center) edge [edgestyle] ([yshift=-0.15cm]line3_gate3.center);
	\draw (line3_gate2) edge[edgestyle] (line3_gate3);
	
	\node[none] (line0_gate46) at (3.25,0.2) {};
	\node[none] (line0_gate47) at (3.7,0.2) {};
	\node[none] (line0_gate48) at (3.25,0.07) {};
	\node[none] (line0_gate49) at (3.7,0.07) {};
	\node[none] (line0_gate50) at (3.25,-0.32) {};
	\node[none] (line0_gate51) at (3.7,-0.32) {};
	\draw (line0_gate46) edge[dashed] (line0_gate47);
	\draw (line0_gate47) edge[edgestyle] (line0_gate6);
	\draw (line0_gate23) edge[edgestyle] (line0_gate46);
	
	\draw (line0_gate48) edge[dashed] (line0_gate49);
	\draw (line0_gate49) edge[edgestyle] (line0_gate7);
	\draw (line0_gate5) edge[edgestyle] (line0_gate48);
	
	\draw (line0_gate50) edge[dashed] (line0_gate51);
	\draw (line0_gate4) edge[edgestyle] (line0_gate50);
	\draw (line0_gate51) edge[edgestyle] (line0_gate37);

	\node[none] (line0_gate60) at (1.35,-0.1) {};
	\node[none] (line0_gate61) at (1.55,-0.1) {};
	\path[] (line0_gate60) node {\tiny  $\cdots$ } (line0_gate61);
	
	\end{tikzpicture}
	\caption{The circuit of $U_{\rm SVE}^{(m)}$, $m=0,\cdots, N_3-1$. The unitary operators $\hat{U}_Q^{(m)}$ and $W_m$ are defined in (\ref{UQhatm}) and (\ref{W_m}) in the proof of Theorem \ref{the QSVE for tensor Ahat}. $d=n_2-1$. $U_{f_m}$ is a unitary operator implemented through oracle with a computable function $f_m(x)=||\hat{A}^{(m)}||_F \cos(x/2).$ }
	\label{Fig.3}
	\end{center}
\end{figure}
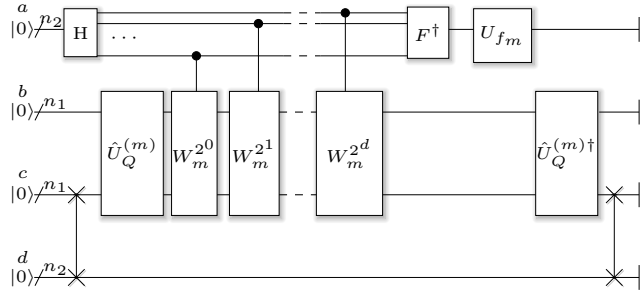

The quantum circuit of Algorithm \ref{alg:Quantum t-svd} is shown in FIG. \ref{Fig.2}, where the block of $U_{\rm SVE}^{(m)}$, $m=0, \cdots, N_3-1,$ is illustrated in FIG. \ref{Fig.3}.

In Step 1, we consider the input state $\ket{\mathcal{A}}$ in (\ref{state tensor A}) and perform the QFT on the third register of this state, obtaining
\begin{equation} \label{second}
\ket{\hat{\mathcal{A}}}=\frac{1}{\sqrt{N_3}}\sum_{m=0}^{N_3-1}\left(\sum_{i,j,k}\omega^{km}\mathcal{A}(i,j,k)\ket{i}^{c}\ket{j}^{d}\right)\ket{m}^{e},
\end{equation}
where $\omega=e^{2\pi\ii/N_3}$.

For every fixed $m$, the unnormalized state
\beq \label{eq:temp1}
\frac{1}{\sqrt{N_3}}\sum_{i,j,k}\omega^{km}\mathcal{A}(i,j,k)\ket{i}\ket{j}
\eeq
 in (\ref{second}) corresponds to the matrix
\begin{align}\label{Ahatm}
\hat{A}^{(m)}=\frac{1}{\sqrt{N_3}}\sum_{i,j,k}\omega^{km}\mathcal{A}(i,j,k)\ket{i}\bra{j},
\end{align}
namely, the $m$-th frontal slice of the tensor $\hat{\mathcal{A}}$.  Normalizing the state in  \eqref{eq:temp1} produces a quantum state
\begin{align} \label{hatAm}
\ket{\hat{A}^{(m)}}&=\frac{1}{\sqrt{N_3}||\hat{A}^{(m)}||_F}\sum_{i,j,k}\omega^{km}\mathcal{A}(i,j,k)\ket{i}^{c}\ket{j}^{d}.
\end{align} Therefore, the state $\ket{\mathcal{A}}$ in (\ref{second}) can be rewritten as
\begin{equation}\label{five}
\ket{\hat{\mathcal{A}}}=\sum_{m=0}^{N_3-1}||\hat{A}^{(m)}||_F\ket{\hat{A}^{(m)}}^{cd}\ket{m}^{e}.
\end{equation}

In Step 2, we design a controlled-$U_{{\rm SVE}}$ operation to estimate  the singular values of $\hat{A}^{(m)}$ parallelly, $ m=0, \cdots, N_3-1$. Denote the procedure of the QSVE on the matrix $\hat{A}^{(m)}$ as $U_{{\rm SVE}}^{(m)}$ and this procedure is unitary \cite{QRS}. Let the svd of $\hat{A}^{(m)}$ in (\ref{Ahatm}) be $\sum_{\ell=0}^{r-1}\hat{\sigma}_{\ell}^{(m)}\hat{u}_\ell^{(m)}\hat{v}_\ell^{(m)\dagger}$, where $\hat{u}_\ell^{(m)}$ and $\hat{v}_\ell^{(m)}$ are the left and right singular vectors corresponding to the singular value $\hat{\sigma}_{\ell}^{(m)}.$  The controlled-$U_{{\rm SVE}}$ is defined as $ \sum_{m=0}^{N_3-1}U_{{\rm SVE}}^{(m)} \otimes \ket{m}\bra{m}$ and when  acting on the input $\ket{\hat{\mathcal{A}}}$ it has the effect of performing the unitary transformation $U_{{\rm SVE}}^{(m)}$ on the state $\ket{\hat{A}^{(m)}}$ for $ m = 0, \cdots, N_3-1$ parallelly. That is,
\begin{align} \label{177}
&\left(\sum_{m=0}^{N_3-1}U_{{\rm SVE}}^{(m)} \otimes \ket{m}\bra{m}\right)\ket{\hat{\mathcal{A}}}   \nonumber \\
=&\sum_{m=0}^{N_3-1}||\hat{A}^{(m)}||_F \left( U_{{\rm SVE}}^{(m)}\ket{\hat{A}^{(m)}}^{(cd)}\right)\ket{m}^{(e)}.
\end{align}
Note that the corresponding input of $U_{{\rm SVE}}^{(m)}$ is $\ket{\hat{A}^{(m)}}$ instead of an arbitrary quantum state commonly used in some quantum svd algorithms \cite{QSVD}. There are mainly three primary reasons for selecting this state as the input. First, after Step 1, the state $\ket{\hat{\mathcal{A}}}$ is the superposition state of $\ket{\hat{A}^{(m)}}$, $m=0,\cdots,N_3-1$. Hence, the operation of $U$ is to perform the $U_{\rm SVE}^{(m)}$ operation on each matrix $\hat{A}^{(m)}$ using the input $\ket{\hat{A}^{(m)}}$ simultaneously, as shown in (\ref{177}). Second, we keep the entire singular information $(\hat{\sigma}_\ell^{(m)}, \hat{u}_\ell^{(m)}, \hat{v}_\ell^{(m)})$ of $\hat{A}^{(m)}$ together, and thus get the quantum svd for tensors similar to the matrix svd formally. The third consideration is that we don't need the information unrelated to the tensor $\mathcal{A}$ (e.g. an arbitrary state) to be involved in the quantum t-svd algorithm.

Next, we focus on the result of $U_{{\rm SVE}}^{(m)}\ket{\hat{A}^{(m)}}$ in (\ref{177}). Following the idea of Remark \ref{modified QSVE remarks}, the input $\ket{\hat{A}^{(m)}}$ can be rewritten in the form $\sum_{\ell}\frac{\hat{\sigma}_\ell^{(m)}}{||\hat{A}^{(m)}||_F}\ket{\hat{u}_\ell^{(m)}} \ket{\hat{v}_\ell^{(m)}}$, where $\frac{\hat{\sigma}_\ell^{(m)}}{||\hat{A}^{(m)}||_F}$ is the scaled singular value of $\frac{\hat{A}^{(m)}}{||\hat{A}^{(m)}||_F}$ because $\ket{\hat{A}^{(m)}}$ is the normalized state corresponding to the matrix $\hat{A}^{(m)}$. Theorem \ref{the QSVE for tensor Ahat} summarizes the above discussions and illustrates the Step 2 of Algorithm \ref{alg:Quantum t-svd}. The proof of Theorem \ref{the QSVE for tensor Ahat} is given in Appendix \ref{appendix Theorem 3.1 }.

\bthm \label{the QSVE for tensor Ahat}
Given every frontal slice of a tensor $\mathcal{A}$ stored in the data structure (Lemma \ref{data structure}), there is a quantum algorithm, denoted by $U_{{\rm SVE}}^{(m)}$, that runs in time $\mathcal{O}({\rm polylog}(N_1N_2)/{\epsilon_{\rm SVE}^{(m)}})$ using the input $\ket{\hat{A}^{(m)}}=\sum_{\ell=0}^{r-1}\frac{\hat{\sigma}_\ell^{(m)}}{||\hat{A}^{(m)}||_F}\ket{\hat{u}_\ell^{(m)}}^{c} \ket{\hat{v}_\ell^{(m)}}^{d}$ and outputs the state
\begin{equation} \label{eq:1 feb3}
\frac{1}{||\hat{A}^{(m)}||_F}\sum_{\ell=0}^{r-1}\hat{\sigma}_\ell^{(m)}\ket{\hat{u}_\ell^{(m)}}^{c} \ket{\hat{v}_\ell^{(m)}}^{d}\ket{\overline{\hat{\sigma}}_\ell^{(m)}}^{a}
\end{equation}
with probability at least $1-1/{\rm poly}(N_2)$, where $(\hat{\sigma}_\ell^{(m)}, \hat{u}_\ell^{(m)}, \hat{v}_\ell^{(m)})$ are the singular pairs of the matrix $\hat{A}^{(m)}$ in (\ref{Ahatm}), and $\epsilon_{\rm SVE}^{(m)}$ is the precision such that $|\overline{\hat{\sigma}}_{\ell}^{(m)}-\hat{\sigma}_{\ell}^{(m)}| \leq \epsilon_{\rm SVE}^{(m)}||\hat{A}^{(m)}||_F$ for all $\ell=0, \cdots,r-1$.
\ethm

Based on Theorem \ref{the QSVE for tensor Ahat}, the state in (\ref{177}) becomes $\ket{\psi}$ in \eqref{eq:psi_jan13}.
Since we have to perform the QSVE on all $\hat{A}^{(m)}$, $m=0,\cdots, N_3-1,$ the running time of Step 2 is $\mathcal{O}(N_3{\rm polylog}(N_1N_2)/{\epsilon_{\rm SVE}})$, where $\epsilon_{\rm SVE}=\min \limits_{0 \leq m \leq N_3-1}\epsilon_{\rm SVE}^{(m)}$.

In Step 3, the inverse QFT is performed on the last register of the state $\ket{\psi}$ in \eqref{eq:psi_jan13} to obtain the final quantum state $\ket{\phi}$ in \eqref{eq:phi_jan13}.

In what follows, we interpret the final quantum state  $\ket{\phi}$ produced by Algorithm  \ref{alg:Quantum t-svd}. First,  according to Algorithm \ref{alg:t-svd} for the classical t-svd,  the singular values of the tensor $\mathcal{A}$ are $\sigma^{(k)}_\ell=\frac{1}{\sqrt{N_3}}\sum_{m=0}^{N_3-1}\omega^{-km}\hat{\sigma}^{(m)}_\ell,$ $\ell=0, \cdots, r-1, \, k=0, \cdots, N_3-1$, where the estimated values of  $\hat{\sigma}^{(m)}_\ell$ are $\overline{\hat{\sigma}}^{(m)}_\ell$ stored in the third register of $\ket{\phi}$, i.e.,  Algorithm  \ref{alg:Quantum t-svd} can produce estimates of the singular values of the original tensor  $\mathcal{A}$. Second, in terms of the circulant matrix ${\rm circ}(\mathcal{A})$ defined in Definition \ref{circtensor},	$\frac{1}{\sqrt{N_3}}\sum_{t=0}^{N_3-1}\omega^{-tm}\ket{t}\ket{\hat{v}_\ell^{(m)}}$ is the right singular vector corresponding to its singular value $\hat{\sigma}_\ell^{(m)}$. Similarly, the corresponding left singular vector is $\frac{1}{\sqrt{N_3}}\sum_{t=0}^{N_3-1}\omega^{-tm}\ket{t}\ket{\hat{u}_\ell^{(m)}}$. Finally, the singular values of $\hat{A}^{(m)}$ have wider applications than the singular values of $\mathcal{A}$. For example, some low-rank tensor completion problems are solved by minimizing the TNN of the tensor, which is defined as the sum of all the singular values of $\hat{A}^{(m)}$ \cite{Novel, Tensor_factorization_for_low-rank}; see Definition \ref{def:tnn}. Moreover, the theoretical minimal error truncation is also based on the singular values of $\hat{A}^{(m)}$; see Lemma \ref{theoretical basis for QRS}. Therefore, in Algorithm \ref{alg:Quantum t-svd}, we estimate the values of $\hat{\sigma}_\ell^{(m)}$, $m=0,\cdots,N_3$, $l=0,\cdots,r-1$, and store them in the third register of the final state for future use.

Our quantum t-svd algorithm can be used as a subroutine of other algorithms, that is, it is suitable for some specific applications where the singular values of $\hat{A}^{(m)}$ are used. For example, in Section \ref{QRS for tensors}, we introduce a quantum recommendation systems algorithm for third order tensors which extracts the singular values of $\hat{A}^{(m)}$ and only keep the greater ones. By doing so, the original tensor $\mathcal{A}$ can be approximated and we can recommend a product to a user according to this reconstructed preference information.

\subsection{Complexity analysis} \label{complexity of quantum t-svd section}

For simplification, we consider the tensor $\mathcal{A} \in \mathbb{R}^{N\times N \times N} $ with the same dimensions on each mode. In Steps 1 and 3, performing the QFT or the inverse QFT parellelly on the third register of the state $\ket{\mathcal{A}}$ achieves the complexity of $\mathcal{O}(({\rm log}N)^2)$, compared with the complexity $\mathcal{O}(N^3{\rm log}N)$ of the FFT performed on $N^2$ tubes of the tensor $\mathcal{A}$ in the classical t-svd algorithm. Moreover, in the classical t-svd, the complexity of performing the matrix svd (Step 2) for all frontal slices of $\mathcal{\hat{A}}$ is $\mathcal{O}(N^4)$. In contrast, in our quantum t-svd algorithm, this step is accelerated by the QSVE whose complexity is $\mathcal{O}({\rm polylog}(N)/{\epsilon_{\rm SVE}})$ on each frontal slice $\hat{A}^{(m)}$, where $\epsilon_{\rm SVE}=\min \limits_{0 \leq m \leq N-1}\epsilon_{\rm SVE}^{(m)}$, $m=0, \cdots, N-1$; hence the Step 2 of our quantum t-svd algorithm achieves the complexity of $\mathcal{O}(N{\rm polylog}(N)/{\epsilon_{\rm SVE}}).$ If we choose $\epsilon_{\rm SVE}=1/{{\rm polylog}(N)}$, the total computational complexity of Algorithm \ref{alg:Quantum t-svd} is $\mathcal{O}(N{\rm polylog}(N))$ which is exponentially faster than the classical t-svd with $\mathcal{O}(N^4)$.

\subsection{Quantum t-svd for order-$p$ tensors} \label{quantum t-svd order-p}
Following a similar procedure, we can extend the quantum t-svd for third-order tensors to order-$p$ tensors easily.

We assume that the quantum state $\ket{\mathcal{A}}$ corresponding to the tensor $\mathcal{A} \in \mathbb{R}^{N_1 \times \cdots \times N_p}$ can be prepared efficiently, where $N_i=2^{n_i}$ with $n_i$ being the number of qubits on the corresponding mode, and
\begin{align}
\ket{\mathcal{A}}=\sum_{i_1=0}^{N_1-1}\cdots \sum_{i_p=0}^{N_p-1}\mathcal{A}(i_1,\cdots, i_p)\ket{i_1,\cdots, i_p}.
\end{align}

Next, we perform the QFT on the third to the $p$-th order of the state $\ket{\mathcal{A}}$, and then use one register $\ket{m}$ to denote $\ket{m_3}\cdots \ket{m_p}$, i.e., $m=m_32^{p-3}+m_42^{p-4}+\cdots +m_p$, $m=0,\cdots, \iota-1,$ $\iota=N_3N_4\cdots N_p$, obtaining
\begin{align}
\ket{\mathcal{\hat{A}}}=&\frac{1}{\sqrt{\iota}}\sum_{m=0}^{\iota-1}\sum_{i_1,\cdots, i_p}\omega^{\sum_{\ell=3}^{p}i_\ell m_\ell} \nonumber \\
& \mathcal{A}(i_1,\cdots, i_p)\ket{i_1,i_2}\ket{m}. \label{order-p 16}
\end{align}

\begin{algorithm}[H]
	\caption{Quantum t-svd for order-$p$ tensors }
	\label{alg:Quantum t-svd for p}
	\hspace*{\algorithmicindent} \textbf{Input:} tensor $\mathcal{A} \in \mathbb{R}^{N_1 \times \cdots \times N_p}$ prepared in a quantum state, precision $\epsilon_{\rm SVE}^{(m)}$, $m=0, \cdots, \iota-1$. \\
	\hspace*{\algorithmicindent} \textbf{Output:} the state $\ket{\phi_p}.$
	\begin{algorithmic}[1]
		\STATE Perform the QFT parallelly from the third to the $p$-th register of quantum state $\ket{\mathcal{A}}$, obtain the state $\ket{\mathcal{\hat{A}}}$.
		\STATE Perform the QSVE for each matrix $\hat{A}^{(m)}$ with precision $\epsilon_{\rm SVE}^{(m)}$ parallelly, $m=0, \cdots, \iota-1$, by using the controlled-$U_{{\rm SVE}}$ acting on the state $\ket{\mathcal{\hat{A}}}$, to obtain the state
\begin{align} \label{eq17: psi_p}
\ket{\psi_p} =\sum_{m=0}^{\iota-1}\left( \sum_{\ell=0}^{r-1}\hat{\sigma}_\ell^{(m)}\ket{\hat{u}_\ell^{(m)}}\ket{\hat{v}_\ell^{(m)}}\ket{\overline{\hat{\sigma}}_\ell^{(m)}}  \right)\ket{m}.
\end{align}
		\STATE Perform the inverse QFT parallelly from the third to the $p$-th register of the above state and output the state
		\begin{align} \label{eq18:phi_p}
			\ket{\phi_p} = & \frac{1}{(\sqrt{N})^{p-2}}\sum_{m_3=0}^{N_3-1} \cdots \sum_{m_p=0}^{N_p-1}\sum_{\ell=0}^{r-1}\hat{\sigma}_\ell^{(m)}\omega^{-\sum_{\ell=3}^{p}i_\ell m_\ell} \\
			&\ket{\hat{u}_\ell^{(m)}}\ket{\hat{v}_\ell^{(m)}}\ket{\overline{\hat{\sigma}}_\ell^{(m)}}\ket{i_3}\cdots \ket{i_p}.
		\end{align}
	\end{algorithmic}
\end{algorithm}

Let the matrix
\begin{align*}
	\hat{A}^{(m)}=\frac{1}{\sqrt{\iota}}\sum_{i_1,i_2}\sum_{i_3,\cdots, i_p}\omega^{\sum_{\ell=3}^{p}i_\ell m_\ell}\mathcal{A}(i_1,\cdots, i_p)\ket{i_1}\bra{i_2}
\end{align*} and perform the QSVE on $\hat{A}^{(m)}$, $m=0,\cdots,\iota-1$, parallelly using the same strategy described in Section \ref{quantum t-svd section}, we can get the state $\ket{\psi_p}$ in (\ref{eq17: psi_p}) after Step 2.

Finally, we recover the $\ket{m_3}\cdots \ket{m_p}$ expression and perform the inverse QFT on the third to the $p$-th register, obtaining the final state $\ket{\phi_p}$ in (\ref{eq18:phi_p}) corresponding to the quantum t-svd of order-$p$ tensor $\mathcal{A}$.

\section{Quantum algorithm for recommendation systems modeled by third-order tensors} \label{QRS for tensors}

In this section, we propose a quantum algorithm for recommendation systems modeled by third-order tensors as an application of the quantum t-svd algorithm developed in Section \ref{quantum t-svd section}. To do this,  Algorithm \ref{alg:Quantum t-svd} has been modified in the following ways. First, the input state to the new algorithm encodes the preference information of user $i$ because we want to output the recommended index for any specific user; see Step 2 of Algorithm \ref{alg:QRS for tensors} for details. Second, after the QFT and the QSVE steps of Algorithm \ref{alg:Quantum t-svd}, we truncate the greater singular values of each frontal slice, and apply the inverse QFT just as Step 3 of Algorithm \ref{alg:Quantum t-svd}. In this way, we get a state which can be proved to be an approximation of the input state. Finally, projection measurement and postselection generate the recommendation index for user $i$.

We will first introduce the notation adopted in this section and then give a brief overview of Algorithm \ref{alg:QRS for tensors}. In Section \ref{subsec:main ideas}, the main ideas and assumptions of the algorithm are summarized.  In Section \ref{Section QRS for tensors}, Algorithm \ref{alg:QRS for tensors} is provided first, followed by the detailed explanation of each step. Theoretical analysis is given in Section \ref{Theoretical analysis} and complexity analysis is conducted in Section \ref{complexity of QRS}. Finally, a quantum algorithm for solving the problem of third-order tensor completion is introduced in Section \ref{Section: another truncate}.

{\it Notation.} The preference information of users is stored in a third-order tensor $\mathcal{T} \in \mathbb{R}^{N\times N\times N}$, called the preference tensor, whose three modes represent user($i$), product($j$) and context($t$) respectively. The tube $\mathcal{T}(i,j,:)$ is regarded as the rating score of the user $i$ for the product $j$ under different contexts. The entry $\mathcal{T}(i,j,t)$ takes value 1 indicating the product $j$ is ``good" for user $i$ in context $t$ and value 0 otherwise. $\mathcal{T}(:,:,m)$ is represented as $T^{(m)}$ (frontal slice). Let tensor $\tilde{\mathcal{T}}$ be the random tensor obtained by sampling from the tensor $\mathcal{T}$ with probability $p$ and $\hat{\tilde{\mathcal{T}}}$ be the tensor obtained by performing the QFT along the third mode of $\tilde{\mathcal{T}}$. The tensor $\hat{\tilde{\mathcal{T}}}_{\geq \sigma}$ denotes the tensor whose $m$-th frontal slice is $\hat{\tilde{T}}_{\geq \sigma^{(m)}}^{(m)}$ formed by truncating the $m$-th frontal slice $\hat{\tilde{T}}^{(m)}$ with a given threshold $\sigma^{(m)}$. $\tilde{\mathcal{T}}_{\geq \sigma}$ denotes the tensor obtained by performing the inverse QFT along the third mode of $\hat{\tilde{\mathcal{T}}}_{\geq \sigma}$. The $i$-th horizontal slice of $\tilde{\mathcal{T}}_{\geq \sigma}$ is $\tilde{\mathcal{T}}_{\geq \sigma}(i,:,:)$. The $i$-th row of a matrix $T$ is represented as $T_i$.

\subsection{Main ideas} \label{subsec:main ideas}

Given a hidden preference tensor $\mathcal{T}$, we will propose Algorithm \ref{alg:QRS for tensors} to recommend a product $j$ to a user $i$ at a certain context $t_0$. The algorithm is inspired by the matrix recommendation methods developed in  \cite{Achlioptas,QRS} and  a tensor reconstruction algorithm \cite{Novel}.
%
%
The main idea is summarized in the following flow chart.
\begin{align*}
&\mathcal{T}(i,:,:) \xrightarrow{\rm sample} \tilde{\mathcal{T}}(i,:,:) \xrightarrow{\rm QFT} \hat{\tilde{\mathcal{T}}}(i,:,:) \xrightarrow{\rm tube} \hat{\tilde{\mathcal{T}}}(i,:,m)  \nonumber \\
& \xrightarrow{\rm approximation}   \hat{\tilde{\mathcal{T}}}_{\geq \sigma}(i,:,m) \xrightarrow{\rm form\, slice}\hat{\tilde{\mathcal{T}}}_{\geq \sigma}(i,:,:) \nonumber \\ & \xrightarrow{\rm iQFT}\tilde{\mathcal{T}}_{\geq \sigma}(i,:,:).
\end{align*}

In Algorithm \ref{alg:QRS for tensors}, we first sample the preference tensor $\mathcal{T}$ with probability $p$, obtaining the tensor $\tilde{\mathcal{T}}$ which represents the preference information that we are able to  collect. That is, $\tilde{\mathcal{T}}_{ijt}=\mathcal{T}_{ijt}/p$ with probability $p$ and $\tilde{\mathcal{T}}_{ijt}=0$ otherwise. Clearly, $\E \left(\tilde{\mathcal{T}}\right)=\mathcal{T}$. Given a state $\ket{\tilde{\mathcal{T}}(i,:,:)}$ representing the user $i$' subsample preference information, we first perform QFT on the last register of the state $\ket{\tilde{\mathcal{T}}(i,:,:)}$, obtaining the state $\ket{\hat{\tilde{\mathcal{T}}}(i,:,:)}$. By performing the QSVE on the $m$-th frontal slice $\hat{\tilde{T}}^{(m)}$ using the input state $\ket{\hat{\tilde{\mathcal{T}}}(i,:,m)}$ and truncating the resulting singular values with threshold $\sigma^{(m)}$, the state $\ket{\hat{\tilde{\mathcal{T}}}_{\geq \sigma}(i,:,m)}$ is obtained. Stacking tubes $\hat{\tilde{\mathcal{T}}}_{\geq \sigma}(i,:,m)$ ($m=0,\ldots,N-1$) yields  the horizontal slice  $\hat{\tilde{\mathcal{T}}}_{\geq \sigma}(i,:,:)$ which can be regarded as an approximation of $\hat{\tilde{\mathcal{T}}}(i,:,:)$. After the inverse QFT on $\hat{\tilde{\mathcal{T}}}_{\geq \sigma}(i,:,:)$, the horizontal slice $\tilde{\mathcal{T}}_{\geq \sigma}(i,:,:)$ is obtained. We can prove that $\tilde{\mathcal{T}}_{\geq \sigma}(i,:,:)$ is an approximation of the original slice $\tilde{\mathcal{T}}(i,:,:)$ in Section \ref{Theoretical analysis}.

\begin{assumption} \label{assumption for QRS}
The following assumptions are used Algorithm \ref{alg:QRS for tensors}.
	\begin{itemize}
		\item[1.]  Each $T^{(m)}$, $m=0,\cdots,N-1,$ has a good rank-$k$ approximation.
		\item[2.]  Every frontal slice of the subsample tensor $\tilde{\mathcal{T}} \in \mathbb{R}^{N\times N\times N}$ is stored in the data structure as mentioned in Lemma \ref{data structure}.
		\item[3.]  For all $i, m=0,\cdots, N-1$, we assume the tubes  $\mathcal{T}(i,:,m)$ satisfty
		\begin{align} \label{123}
		\frac{1}{1+\gamma}\frac{||\mathcal{T}||_F^2}{N^2} \leq ||\mathcal{T}(i,:,m)||_2^2 \leq (1+\gamma)\frac{||\mathcal{T}||_F^2}{N^2}
		\end{align}
for a given $\gamma >0$.		
	\end{itemize}
\end{assumption}
The first assumption is reasonable because most of users belong to a small number of types, and the third assumption indicates that users in the preference tensor $\mathcal{T}$ are all typical users. In other words, the number of preferred products of users is close to the average in any context $m$. These assumptions are also adopted in Kerenidis and Prakash's work \cite{QRS} for matrices, where they give detailed explanation to justify.

\subsection{Quantum algorithm for recommendation systems modeled by third-order tensors}\label{Section QRS for tensors}

Algorithm \ref{alg:QRS for tensors} is a quantum algorithm that,  given the dynamic preference tensor $\mathcal{T}$, the sampling probability $p$, the assumed low rank $k$, the threshold $\sigma^{(m)}$, and the precision $\epsilon_{\rm SVE}^{(m)}$ for QSVE on each $\hat{\tilde{T}}^{(m)}$, outputs the state corresponding to the approximation of the $i$-th horizontal slice $\mathcal{T}(i,:,:)$.  The algorithm is stated below.

Algorithm \ref{alg:QRS for tensors}  is given below, whose circuit is shown in FIGs. \ref{Figure 4} and \ref{Fig.5}.

\begin{algorithm}[H]
	\caption{Quantum algorithm for recommendation systems modeled by third-order tensors}
	\label{alg:QRS for tensors}
	\hspace*{\algorithmicindent} \textbf{Input:} a user index $i$, the state $\ket{\tilde{\mathcal{T}}(i,:,:)}$ corresponding to the preference information of user $i$ , precision $\epsilon_{\rm SVE}^{(m)}$, the truncation threshold $\sigma^{(m)}$,  $m=0, \cdots, N-1$,  and a context $t_0$. \\
	\hspace*{\algorithmicindent} \textbf{Output:} the recommended index $j$ for the user $i$ at the context $t_0.$
	\begin{algorithmic}[1]
		\STATE Perform the QFT on the last register of the input state $\ket{\tilde{\mathcal{T}}(i,:,:)}$, to obtain $\ket{\hat{\tilde{\mathcal{T}}}(i,:,:)}$ in (\ref{Step 1}).
		
		\STATE Perform the QSVE on the matrix $\hat{\tilde{T}}^{(m)}$ parallelly, using the input $\ket{\hat{\tilde{\mathcal{T}}}(i,:,:)}$ with precision $\epsilon_{\rm SVE}^{(m)}$, $m=0, \cdots, N-1,$ to get the state $\ket{\xi_1}$ defined in (\ref{Step 2}).
		
		\STATE Add an ancilla qubit $\ket{0}^{a}$ and apply a unitary transformation $V$ on the registers $b$ and $a$, controlled by the register $e$ (see FIG. \ref{Figure 4}), to obtain $\ket{\xi_2}$ in (\ref{Step 3}).

		\STATE Apply the inverse QSVE and discard the register $c$, to get $\ket{\xi_3}$ in (\ref{Step 4}).

		\STATE Measure the ancilla register $a$ in the computational basis and postselect the outcome $\ket{0}$, then delete the register $a$, to obtain $\ket{\xi_4}$ in (\ref{Step 5}).

		\STATE Perform the inverse QFT on the register $e$, to get $\ket{\xi_5}$ in (\ref{final state 16}).

		\STATE Measure the register $e$ in the computational basis and postselect the outcome $\ket{t_0}$. Then measure the register $d$ in the computational basis to get the index $j$.
		
	\end{algorithmic}
\end{algorithm}
Next, we explain each step in detail.

The dynamic preference tensor $\mathcal{T} \in \mathbb{R}^{N\times N\times N}$ can be interpreted as the preference matrix $\mathcal{T}(:,:,t)$ evolving over the context $t$. It is reasonable to believe that the tubes $\mathcal{T}(i,:,t), \cdots, \mathcal{T}(i,:,N-1)$ are related to each other because the preference of the same user $i$ in different contexts is mutually influenced. Considering these relations, we merge tubes in the same horizontal slice together through the QFT after getting the subsample tensor $\tilde{\mathcal{T}}$. In other words,  in Step 1, the QFT is performed on the last register of the input state
\begin{equation} \label{eq1}
\ket{\tilde{\mathcal{T}}(i,:,:)}=\frac{1}{||\tilde{\mathcal{T}}(i,:,:)||_F}\sum_{j,t=0}^{N-1}\tilde{\mathcal{T}}(i,j,t)\ket{j}^{d}\ket{t}^{e}
\end{equation}
to get
\begin{align}
&\ket{\hat{\tilde{\mathcal{T}}}(i,:,:)} \nonumber \\
=&\frac{1}{||\tilde{\mathcal{T}}(i,:,:)||_F}\sum_{m=0}^{N-1}||\hat{\tilde{\mathcal{T}}}(i,:,m)||_2\ket{\hat{\tilde{\mathcal{T}}}(i,:,m)}^{d}\ket{m}^{e}, \label{Step 1}
\end{align}
where $\omega=e^{2\pi\ii/N}$ and
\begin{align} \label{110}
&\ket{\hat{\tilde{\mathcal{T}}}(i,:,m)}
 \nonumber \\
  =&\frac{1}{\sqrt{N}||\hat{\tilde{\mathcal{T}}}(i,:,m)||_2}
\sum_{j,t=0}^{N-1}\omega^{tm}\tilde{\mathcal{T}}(i,j,t)\ket{j}.
\end{align} Note that $||\hat{\tilde{\mathcal{T}}}(i,:,:)||_F=||\tilde{\mathcal{T}}(i,:,:)||_F,$ since the Frobenius norm of $\hat{\tilde{\mathcal{T}}}(i,:,:)$ does not change when performing the Fourier transform.
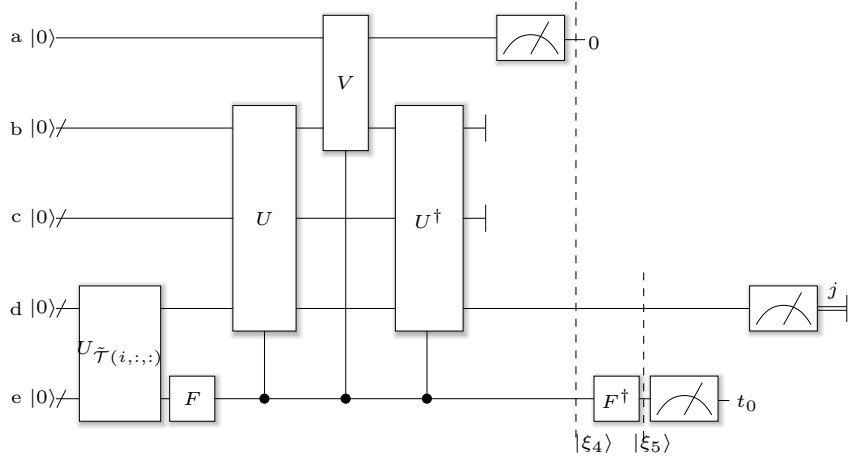
\begin{figure}[H]
\begin{center}	
	\begin{tikzpicture}[scale=1.2, transform shape]
	
	\tikzstyle{basicshadow}=[blur shadow={shadow blur steps=8, shadow xshift=0.7pt, shadow yshift=-0.7pt, shadow scale=1.02}]\tikzstyle{basic}=[draw,fill=white,basicshadow]
	\tikzstyle{operator}=[basic,minimum size=1.5em]
	\tikzstyle{phase}=[fill=black,shape=circle,minimum size=0.1cm,inner sep=0pt,outer sep=0pt,draw=black]
	\tikzstyle{none}=[inner sep=0pt,outer sep=-.5pt,minimum height=0.5cm+1pt]
	\tikzstyle{measure}=[operator,inner sep=0pt,minimum height=0.5cm, minimum width=0.75cm]
	\tikzstyle{xstyle}=[circle,basic,minimum height=0.35cm,minimum width=0.35cm,inner sep=-1pt,very thin]
	\tikzset{
		shadowed/.style={preaction={transform canvas={shift={(0.5pt,-0.5pt)}}, draw=gray, opacity=0.4}},
	}
	\tikzstyle{swapstyle}=[inner sep=-1pt, outer sep=-1pt, minimum width=0pt]
	\tikzstyle{edgestyle}=[very thin]
	
	\node[none] (line0_gate0) at (0.1,-0) {\tiny $\Ket{0}$};
	\node[none] (line1_gate0) at (0.1,-1) {\tiny $\Ket{0}$};
	\node[none] (line2_gate0) at (0.1,-2) {\tiny $\Ket{0}$};
	\node[none] (line3_gate0) at (0.1,-3) {\tiny $\Ket{0}$};
	\node[none] (line4_gate0) at (0.1,-4) {\tiny $\Ket{0}$};
	\node[none] (line3_gate1) at (0.5,-3) {};
	\node[none,minimum height=0.5cm,outer sep=0] (line3_gate2) at (0.9,-3) {};
	\node[none] (line3_gate3) at (1.4,-3) {};
	\node[none] (line4_gate1) at (0.5,-4) {};
	\node[none,minimum height=0.5cm,outer sep=0] (line4_gate2) at (0.9,-4) {};
	\node[none] (line4_gate3) at (1.4,-4) {};
	\draw[operator,edgestyle,outer sep=0.5cm] ([yshift=0.25cm]line3_gate1) rectangle ([yshift=-0.25cm]line4_gate3) node[pos=.5] {\tiny $U_{\tilde{\mathcal{T}}(i,:,:)}$};
	\draw (line3_gate0) edge[edgestyle] (line3_gate1);
	\draw (line4_gate0) edge[edgestyle] (line4_gate1);
	\node[none] (line4_gate4) at (1.5,-4) {};
	\node[none,minimum height=0.5cm,outer sep=0] (line4_gate5) at (1.75,-4) {};
	\node[none] (line4_gate6) at (2.0,-4) {};
	\draw[operator,edgestyle,outer sep=0.5cm] ([yshift=0.25cm]line4_gate4) rectangle ([yshift=-0.25cm]line4_gate6) node[pos=.5] {\tiny $F$};
	\draw (line4_gate3) edge[edgestyle] (line4_gate4);
	\node[none] (line1_gate1) at (2.2000000000000003,-1) {};
	\node[none,minimum height=0.5cm,outer sep=0] (line1_gate2) at (2.5500000000000003,-1) {};
	\node[none] (line1_gate3) at (2.9000000000000003,-1) {};
	\node[none] (line2_gate1) at (2.2000000000000003,-2) {};
	\node[none,minimum height=0.5cm,outer sep=0] (line2_gate2) at (2.5500000000000003,-2) {};
	\node[none] (line2_gate3) at (2.9000000000000003,-2) {};
	\node[none] (line3_gate4) at (2.2000000000000003,-3) {};
	\node[none,minimum height=0.5cm,outer sep=0] (line3_gate5) at (2.5500000000000003,-3) {};
	\node[none] (line3_gate6) at (2.9000000000000003,-3) {};
	\draw[operator,edgestyle,outer sep=0.5cm] ([yshift=0.25cm]line1_gate1) rectangle ([yshift=-0.25cm]line3_gate6) node[pos=.5] {\tiny $U$};
	\node[phase] (line4_gate7) at (2.5500000000000003,-4) {};
	\draw (line4_gate7) edge[edgestyle] (line3_gate5);
	\draw (line1_gate0) edge[edgestyle] (line1_gate1);
	\draw (line2_gate0) edge[edgestyle] (line2_gate1);
	\draw (line3_gate3) edge[edgestyle] (line3_gate4);
	\draw (line4_gate6) edge[edgestyle] (line4_gate7);
	\node[none] (line0_gate1) at (3.200000000000003,-0) {};
	\node[none,minimum height=0.5cm,outer sep=0] (line0_gate2) at (3.4500000000000003,-0) {};
	\node[none] (line0_gate3) at (3.7000000000000003,-0) {};
	
	\node[none] (line1_gate60) at (3.2000000000000003,-1) {};
	\node[none,minimum height=0.5cm,outer sep=0] (line1_gate61) at (3.4500000000000003,-1) {};
	\node[none] (line1_gate62) at (3.7000000000000003,-1) {};
	\draw[operator,edgestyle,outer sep=0.5cm] ([yshift=0.25cm]line0_gate1) rectangle ([yshift=-0.25cm]line1_gate62) node[pos=.5] {\tiny $V$};
	
	\node[phase] (line4_gate169) at (3.4500000000000003,-4) {};
	\draw (line4_gate169) edge[edgestyle] (line1_gate61);
	\draw (line0_gate0) edge[edgestyle] (line0_gate1);
	\draw (line1_gate3) edge[edgestyle] (line1_gate60);
	\node[measure,edgestyle] (line0_gate4) at (5.5,-0) {};
	\draw[edgestyle] ([yshift=-0.18cm,xshift=0.07500000000000001cm]line0_gate4.west) to [out=60,in=180] ([yshift=0.035cm]line0_gate4.center) to [out=0, in=120] ([yshift=-0.18cm,xshift=-0.07500000000000001cm]line0_gate4.east);
	\draw[edgestyle] ([yshift=-0.18cm]line0_gate4.center) to ([yshift=-0.07500000000000001cm,xshift=-0.18cm]line0_gate4.north east);
	\draw (line0_gate3) edge[edgestyle] (line0_gate4);
	\node[none] (line0_gate5) at (6.1,-0) {};
	\node[none] (line0_gate6) at (6.2,-0.05) {\tiny $0$};
	\node[none] (line0_gate10) at (6,0.4) {};
	\node[none] (line4_gate20) at (6,-4.5) {};
	\draw (line0_gate10) edge[dashed] (line4_gate20);
	\node[none] (line4_gate21) at (6.2,-4.5) {\tiny $\Ket{\xi_4}$};
	
	\node[none] (line3_gate20) at (6.75,-2.6) {};
	\node[none] (line4_gate28) at (6.75,-4.5) {};
	\draw (line3_gate20) edge[dashed] (line4_gate28);
	\node[none] (line4_gate55) at (6.85,-4.5) {\tiny $\Ket{\xi_5}$};

	\draw ([yshift=-0.02cm]line0_gate4.east) edge[edgestyle] ([yshift=-0.02cm]line0_gate5.west);
	\node[none] (line1_gate5) at (4.0,-1) {};
	\node[none,minimum height=0.5cm,outer sep=0] (line1_gate6) at (4.35,-1) {};
	\node[none] (line1_gate7) at (4.75,-1) {};
	\node[none] (line2_gate4) at (4.0,-2) {};
	\node[none,minimum height=0.5cm,outer sep=0] (line2_gate5) at (4.35,-2) {};
	\node[none] (line2_gate6) at (4.75,-2) {};
	\node[none] (line3_gate7) at (4.0,-3) {};
	\node[none,minimum height=0.5cm,outer sep=0] (line3_gate8) at (4.35,-3) {};
	\node[none] (line3_gate9) at (4.75,-3) {};
	\draw[operator,edgestyle,outer sep=0.5cm] ([yshift=0.25cm]line1_gate5) rectangle ([yshift=-0.25cm]line3_gate9) node[pos=.5] {\tiny $U^{\dagger}$};
	\node[phase] (line4_gate8) at (4.35,-4) {};
	\draw (line4_gate8) edge[edgestyle] (line3_gate8);
	\draw (line1_gate62) edge[edgestyle] (line1_gate5);
	\draw (line2_gate3) edge[edgestyle] (line2_gate4);
	\draw (line3_gate6) edge[edgestyle] (line3_gate7);
	\draw (line4_gate7) edge[edgestyle] (line4_gate8);
	\node[none] (line1_gate8) at (4.999999999999999,-1) {};
	\draw ([yshift=0.15cm]line1_gate8.center) edge [edgestyle] ([yshift=-0.15cm]line1_gate8.center);
	\draw (line1_gate7) edge[edgestyle] (line1_gate8);
	\node[none] (line2_gate7) at (4.999999999999999,-2) {};
	\draw ([yshift=0.15cm]line2_gate7.center) edge [edgestyle] ([yshift=-0.15cm]line2_gate7.center);
	\draw (line2_gate6) edge[edgestyle] (line2_gate7);
	\node[measure,edgestyle] (line3_gate10) at (8.3,-3) {};
	\draw[edgestyle] ([yshift=-0.18cm,xshift=0.07500000000000001cm]line3_gate10.west) to [out=60,in=180] ([yshift=0.035cm]line3_gate10.center) to [out=0, in=120] ([yshift=-0.18cm,xshift=-0.07500000000000001cm]line3_gate10.east);
	\draw[edgestyle] ([yshift=-0.18cm]line3_gate10.center) to ([yshift=-0.07500000000000001cm,xshift=-0.18cm]line3_gate10.north east);
	\draw (line3_gate9) edge[edgestyle] (line3_gate10);
	\node[none] (line3_gate11) at (9,-3) {};
	\draw ([yshift=0.15cm]line3_gate11.center) edge [edgestyle] ([yshift=-0.15cm]line3_gate11.center);
	\draw ([yshift=0.02cm]line3_gate10.east) edge[edgestyle] ([yshift=0.02cm]line3_gate11.west);
	\draw ([yshift=-0.02cm]line3_gate10.east) edge[edgestyle] ([yshift=-0.02cm]line3_gate11.west);
	\node[none] (line4_gate9) at (6.2,-4) {};
	\node[none,minimum height=0.5cm,outer sep=0] (line4_gate10) at (6.45,-4) {};
	\node[none] (line4_gate11) at (6.7,-4) {};
	\draw[operator,edgestyle,outer sep=0.5cm] ([yshift=0.25cm]line4_gate9) rectangle ([yshift=-0.25cm]line4_gate11) node[pos=.5] {\tiny $F^{\dagger}$};
	\draw (line4_gate8) edge[edgestyle] (line4_gate9);
	\node[measure,edgestyle] (line4_gate12) at (7.2,-4) {};
	\draw[edgestyle] ([yshift=-0.18cm,xshift=0.07500000000000001cm]line4_gate12.west) to [out=60,in=180] ([yshift=0.035cm]line4_gate12.center) to [out=0, in=120] ([yshift=-0.18cm,xshift=-0.07500000000000001cm]line4_gate12.east);
	\draw[edgestyle] ([yshift=-0.18cm]line4_gate12.center) to ([yshift=-0.07500000000000001cm,xshift=-0.18cm]line4_gate12.north east);
	\draw (line4_gate11) edge[edgestyle] (line4_gate12);
	\node[none] (line4_gate13) at (7.7,-4) {};
	\node[none] (line4_gate15) at (7.9,-4.05) {\tiny $t_0$};

	\node[none] (line3_gate13) at (8.85,-2.8) {\tiny $j$};
	\draw ([yshift=-0.02cm]line4_gate12.east) edge[edgestyle] ([yshift=-0.02cm]line4_gate13.west);
	
	\node[none] (line1_gate40) at (0.25,-1.1){};
	\node[none] (line1_gate41) at (0.35,-0.9){};
	\draw (line1_gate40) edge[edgestyle] (line1_gate41);
	
	\node[none] (line2_gate40) at (0.25,-2.1){};
	\node[none] (line2_gate41) at (0.35,-1.9){};
	\draw (line2_gate40) edge[edgestyle] (line2_gate41);
	
	\node[none] (line3_gate40) at (0.25,-3.1){};
	\node[none] (line3_gate41) at (0.35,-2.9){};
	\draw (line3_gate40) edge[edgestyle] (line3_gate41);
	
	\node[none] (line4_gate40) at (0.25,-4.1){};
	\node[none] (line4_gate41) at (0.35,-3.9){};
	\draw (line4_gate40) edge[edgestyle] (line4_gate41);
	
	\node[none] (line0_gate45) at (-0.2,0) {\tiny a};
	\node[none] (line1_gate45) at (-0.2,-1) {\tiny b};
	\node[none] (line2_gate45) at (-0.2,-2) {\tiny c};
	\node[none] (line3_gate45) at (-0.2,-3) {\tiny d};
	\node[none] (line4_gate45) at (-0.2,-4) {\tiny e};
	
	\end{tikzpicture}
	\caption{The circuit of Algorithm \ref{alg:QRS for tensors}. $U_{\tilde{\mathcal{T}}(i,:,:)}$ is the unitary operator for preparing the initial state $\ket{\tilde{\mathcal{T}}(i,:,:)}$. The unitary operator $U=\sum_{m=0}^{N-1}U_{{\rm SVE}}^{(m)} \otimes \ket{m}\bra{m}$ and the block of $U_{\rm SVE}^{(m)}$ is shown in FIG. \ref{Fig.5}. After measuring the first register in the computational basis, we postselect the outcome $\ket{0}^a,$ getting $\ket{\xi_4}$. } \label{Figure 4}
	\end{center}
\end{figure}

\begin{figure}[H]
\begin{center}
	\begin{tikzpicture}[scale=1.5, transform shape]
	\tikzstyle{basicshadow}=[blur shadow={shadow blur steps=8, shadow xshift=0.7pt, shadow yshift=-0.7pt, shadow scale=1.02}]\tikzstyle{basic}=[draw,fill=white,basicshadow]
	\tikzstyle{operator}=[basic,minimum size=1.5em]
	\tikzstyle{phase}=[fill=black,shape=circle,minimum size=0.1cm,inner sep=0pt,outer sep=0pt,draw=black]
	\tikzstyle{none}=[inner sep=0pt,outer sep=-.5pt,minimum height=0.5cm+1pt]
	\tikzstyle{measure}=[operator,inner sep=0pt,minimum height=0.5cm, minimum width=0.75cm]
	\tikzstyle{xstyle}=[circle,basic,minimum height=0.35cm,minimum width=0.35cm,inner sep=-1pt,very thin]
	\tikzset{
		shadowed/.style={preaction={transform canvas={shift={(0.5pt,-0.5pt)}}, draw=gray, opacity=0.4}},
	}
	\tikzstyle{swapstyle}=[inner sep=-1pt, outer sep=-1pt, minimum width=0pt]
	\tikzstyle{edgestyle}=[very thin]
	
	\node[none] (line0_gate0) at (0.1,-0) {\tiny $\Ket{0}$};
	\node[none] (line0_gate1) at (0.6,-0) {};
	\node[none,minimum height=0.5cm,outer sep=0] (line0_gate2) at (0.85,-0) {};
	\node[none] (line0_gate3) at (1,0.07) {};
	\draw[operator,edgestyle,outer sep=0.5cm] ([yshift=0.25cm]line0_gate1) rectangle ([yshift=-0.4cm]line0_gate3) node[pos=.5] {\tiny $H$};
	\draw (line0_gate0) edge[edgestyle] (line0_gate1);
	\node[none] (line1_gate0) at (0.1,-1) {\tiny $\Ket{0}$};
	\node[none] (line2_gate0) at (0.1,-2){};
	\node[none] (line2_gate2) at (1.05,-2) {};

	\draw (line2_gate0) edge[edgestyle] (line2_gate2);
	
	\node[none] (line1_gate1) at (1.05,-1) {};
	\node[none,minimum height=0.5cm,outer sep=0] (line1_gate2) at (1.4,-1) {};
	\node[none] (line1_gate3) at (1.7,-1) {};
	
	\node[none,minimum height=0.5cm,outer sep=0] (line2_gate3) at (1.4,-2) {};
	\node[none] (line2_gate4) at (1.7,-2) {};
	\draw[operator,edgestyle,outer sep=0.5cm] ([yshift=0.25cm]line1_gate1) rectangle ([yshift=-0.25cm]line2_gate4) node[pos=.5] {\tiny $\hat{U}_Q^{(m)}$};
	\draw (line1_gate0) edge[edgestyle] (line1_gate1);
	
	\node[none] (line1_gate4) at (1.900000000000002,-1) {};
	\node[none,minimum height=0.5cm,outer sep=0] (line1_gate5) at (2.2,-1) {};
	\node[none] (line1_gate6) at (2.45,-1) {};
	\node[none] (line2_gate5) at (1.900000000000002,-2) {};
	\node[none,minimum height=0.5cm,outer sep=0] (line2_gate6) at (2.2,-2) {};
	\node[none] (line2_gate7) at (2.45,-2) {};
	\draw[operator,edgestyle,outer sep=0.5cm] ([yshift=0.25cm]line1_gate4) rectangle ([yshift=-0.25cm]line2_gate7) node[pos=.5] {\tiny $W_m^{2^0}$};
	\node[phase] (line0_gate4) at (2.2,-0.32) {};
	\draw (line0_gate4) edge[edgestyle] (line1_gate5);
	\draw (line1_gate3) edge[edgestyle] (line1_gate4);
	\draw (line2_gate4) edge[edgestyle] (line2_gate5);

	\node[none] (line0_gate30) at (0.25,-0.1){};
	\node[none] (line0_gate31) at (0.35,0.1){};
	\draw (line0_gate30) edge[edgestyle] (line0_gate31);
	
	\node[none] (line1_gate30) at (0.25,-1.1){};
	\node[none] (line1_gate31) at (0.35,-0.9){};
	\draw (line1_gate30) edge[edgestyle] (line1_gate31);
	
	\node[none] (line2_gate30) at (0.25,-2.1){};
	\node[none] (line2_gate31) at (0.35,-1.9){};
	\draw (line2_gate30) edge[edgestyle] (line2_gate31);

	\node[none] (line0_gate23) at (1,0.2) {};
	\node[phase] (line0_gate6) at (4.0,0.2) {};
	
	\node[none] (line0_gate27) at (4.75,0.2) {};
	\draw (line0_gate6) edge[edgestyle] (line0_gate27);
	\node[phase] (line0_gate5) at (2.95,0.07) {};
	\node[none] (line0_gate7) at (4.750000000000001,0.07) {};
	\draw (line0_gate3) edge[edgestyle] (line0_gate5);

	\node[none] (line0_gate18) at (1,-0.32) {};
	\node[none] (line0_gate37) at (4.75,-0.32) {};
	
	\draw (line0_gate4) edge[edgestyle] (line0_gate18);
	
	\node[none] (line1_gate7) at (2.6,-1) {};
	\node[none,minimum height=0.5cm,outer sep=0] (line1_gate8) at (2.95,-1) {};
	\node[none] (line1_gate9) at (3.2,-1) {};
	\node[none] (line1_gate10) at (3.6500000000000004,-1) {};
	\node[none,minimum height=0.5cm,outer sep=0] (line1_gate11) at (4.0,-1) {};
	\node[none] (line1_gate12) at (4.45,-1) {};

	\node[none] (line2_gate8) at (2.6,-2) {};
	\node[none,minimum height=0.5cm,outer sep=0] (line2_gate9) at (2.95,-2) {};
	\node[none] (line2_gate10) at (3.2,-2) {};
	\draw[operator,edgestyle,outer sep=0.5cm] ([yshift=0.25cm]line1_gate7) rectangle ([yshift=-0.25cm]line2_gate10) node[pos=.5] {\tiny $W_m^{2^1}$};
	\draw (line1_gate6) edge[edgestyle] (line1_gate7);
	\draw (line2_gate7) edge[edgestyle] (line2_gate8);
	\node[none] (line2_gate11) at (3.6500000000000004,-2) {};
	\draw (line1_gate9) edge[dashed] (line1_gate10);
	\draw (line2_gate10) edge[dashed] (line2_gate11);
	\draw (line0_gate5) edge[edgestyle] (line1_gate8);

	\node[none,minimum height=0.5cm,outer sep=0] (line2_gate12) at (4.0,-2) {};
	\node[none] (line2_gate13) at (4.45,-2) {};
	\draw[operator,edgestyle,outer sep=0.5cm] ([yshift=0.25cm]line1_gate10) rectangle ([yshift=-0.25cm]line2_gate13) node[pos=.5] {\tiny $W_m^{2^d}$};
	\draw (line0_gate6) edge[edgestyle] (line1_gate11);
	
	\node[none] (line0_gate45) at (0.45,0.1) {};
	\node[none] (line1_gate45) at (0.5,-0.9) {};
	\node[none] (line2_gate45) at (0.5,-1.9) {};

	\node[none,minimum height=0.5cm,outer sep=0] (line0_gate8) at (5.000000000000001,-0) {};
	\node[none] (line0_gate9) at (5.250000000000001,-0) {};
	\draw[operator,edgestyle,outer sep=0.5cm] ([yshift=0.2cm]line0_gate7) rectangle ([yshift=-0.38cm]line0_gate9) node[pos=.5] {\tiny $F^{\dagger}$};
	
	\node[none] (line0_gate10) at (5.550000000000001,-0) {};
	\node[none,minimum height=0.5cm,outer sep=0] (line0_gate11) at (5.900000000000001,-0) {};
	\node[none] (line0_gate12) at (6.250000000000001,-0) {};
	\draw[operator,edgestyle,outer sep=0.5cm] ([yshift=0.25cm]line0_gate10) rectangle ([yshift=-0.4cm]line0_gate12) node[pos=.5] {\tiny $U_{f_m}$};
	\draw (line0_gate9) edge[edgestyle] (line0_gate10);
	\node[none] (line0_gate13) at (7.150000000000001,-0) {};
	\draw ([yshift=0.15cm]line0_gate13.center) edge [edgestyle] ([yshift=-0.15cm]line0_gate13.center);
	\draw (line0_gate12) edge[edgestyle] (line0_gate13);
	\node[none] (line1_gate13) at (6.3,-1) {};
	\node[none,minimum height=0.5cm,outer sep=0] (line1_gate14) at (6.65,-1) {};
	\node[none] (line1_gate15) at (7,-1) {};
	\node[none] (line2_gate14) at (6.3,-2) {};
	\node[none,minimum height=0.5cm,outer sep=0] (line2_gate15) at (6.65,-2) {};
	\node[none] (line2_gate16) at (7,-2) {};
	\draw[operator,edgestyle,outer sep=0.5cm] ([yshift=0.25cm]line1_gate13) rectangle ([yshift=-0.25cm]line2_gate16) node[pos=.5] {\tiny $\hat{U}_Q^{(m)\dagger}$};
	\draw (line1_gate12) edge[edgestyle] (line1_gate13);
	\draw (line2_gate13) edge[edgestyle] (line2_gate14);
	\node[none] (line1_gate16) at (7.15,-1) {};
	\draw (line1_gate15) edge[edgestyle] (line1_gate16);
	\draw ([yshift=0.15cm]line1_gate16.center) edge [edgestyle] ([yshift=-0.15cm]line1_gate16.center);
	
	\node[none] (line2_gate18) at (7.15,-2) {};
	\draw ([yshift=0.15cm]line2_gate18.center) edge [edgestyle] ([yshift=-0.15cm]line2_gate18.center);
	\draw (line2_gate16) edge[edgestyle] (line2_gate18);

	\node[none] (line2_gate70) at (6.7,-2) {};
	
	\node[none] (line0_gate46) at (3.25,0.2) {};
	\node[none] (line0_gate47) at (3.7,0.2) {};
	\node[none] (line0_gate48) at (3.25,0.07) {};
	\node[none] (line0_gate49) at (3.7,0.07) {};
	\node[none] (line0_gate50) at (3.25,-0.32) {};
	\node[none] (line0_gate51) at (3.7,-0.32) {};
	\draw (line0_gate46) edge[dashed] (line0_gate47);
	\draw (line0_gate47) edge[edgestyle] (line0_gate6);
	\draw (line0_gate23) edge[edgestyle] (line0_gate46);
	
	\draw (line0_gate48) edge[dashed] (line0_gate49);
	\draw (line0_gate49) edge[edgestyle] (line0_gate7);
	\draw (line0_gate5) edge[edgestyle] (line0_gate48);
	
	\draw (line0_gate50) edge[dashed] (line0_gate51);
	\draw (line0_gate4) edge[edgestyle] (line0_gate50);
	\draw (line0_gate51) edge[edgestyle] (line0_gate37);

	\node[none] (line0_gate60) at (1.35,-0.1) {};
	\node[none] (line0_gate61) at (1.55,-0.1) {};
	\path[] (line0_gate60) node {\tiny  $\cdots$ } (line0_gate61);
	
	\node[none] (line0_gate45) at (-0.2,0) {\tiny b};
	\node[none] (line1_gate45) at (-0.2,-1) {\tiny c};
	\node[none] (line2_gate45) at (-0.2,-2) {\tiny d};
	
	\end{tikzpicture}
	\caption{The implementation of $U_{\rm SVE}^{(m)}$.} \label{Fig.5}
	\end{center}
\end{figure}
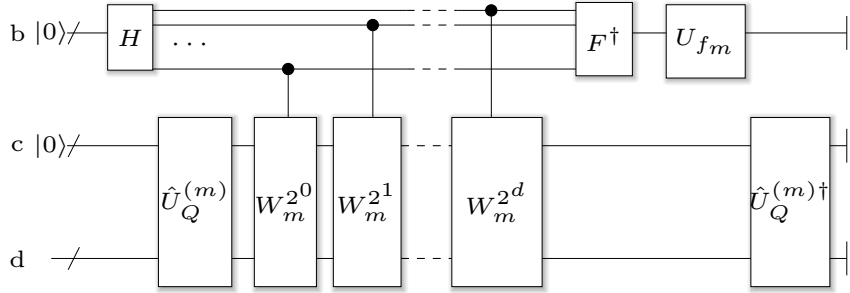

In Step 2, a unitary operator $U=\sum_{m=0}^{N-1}U_{{\rm SVE}}^{(m)} \otimes \ket{m}\bra{m}$, given in FIG. \ref{Figure 4}, is performed on the state $\ket{\hat{\tilde{\mathcal{T}}}(i,:,:)}$ from Step 1. Here, $U_{{\rm SVE}}^{(m)}$ denotes the QSVE procedure for the matrix $\hat{\tilde{T}}^{(m)}$ with the input $\ket{\hat{\tilde{\mathcal{T}}}(i,:,m)}$. This step borrows the idea of Step 2 of Algorithm \ref{alg:Quantum t-svd}. Based on Theorem \ref{the QSVE for tensor Ahat} and the analysis of Algorithm \ref{alg:Quantum t-svd} and Lemma \ref{QSVE}, Step 2 can be expressed as the following transformation:
\begin{align}
& U\ket{\hat{\tilde{\mathcal{T}}}(i,:,:)}=\frac{1}{||\tilde{\mathcal{T}}(i,:,:)||_F} \nonumber \\
&\sum_{m=0}^{N-1}
||\hat{\tilde{\mathcal{T}}}(i,:,m)||_2 \left( U_{{\rm SVE}}^{(m)} \ket{\hat{\tilde{\mathcal{T}}}(i,:,m)}^{d} \right)\ket{m}^{e}. \label{eq29}
\end{align}
Then we express $\ket{\hat{\tilde{\mathcal{T}}}(i,:,m)}$ under the basis of $\hat{v}_j^{(m)}, j=0, \cdots, N-1$, i.e., \begin{align} \label{eq:25}
	\ket{\hat{\tilde{\mathcal{T}}}(i,:,m)}=
	\sum_{j=0}^{N-1} \beta_{j}^{(im)} \ket{\hat{v}_j^{(m)}},
\end{align} where $\sum\limits_{j=0}^{N-1}\hat{\sigma}_j^{(m)}\hat{u}_j^{(m)}\hat{v}_j^{{(m)}\dagger}$ is the svd of $\hat{\tilde{T}}^{(m)}$. According to Lemma \ref{QSVE}, (\ref{eq29}) becomes
\begin{align}
&\frac{1}{||\tilde{\mathcal{T}}(i,:,:)||_F}\sum_{m,j}||\hat{\tilde{\mathcal{T}}}(i,:,m)||_2 \beta_{j}^{(im)} \ket{\hat{v}_j^{(m)}}^{d}\ket{\overline{\hat{\sigma}}_j^{(m)}}^{b}\ket{m}^{e} \nonumber \\
& \triangleq \ket{\xi_1},   \label{Step 2}
\end{align} where $\overline{\hat{\sigma}}_j^{(m)}$ is an estimate of $\hat{\sigma}_j^{(m)}$ such that $|\overline{\hat{\sigma}}_j^{(m)}-\hat{\sigma}_j^{(m)}| \leq \epsilon_{\rm SVE}^{(m)}||\hat{\tilde{T}}^{(m)}||_F$.

In Steps 3-5, our goal is to project each tube $\hat{\tilde{\mathcal{T}}}(i,:,m)$ onto the subspace $\left(\hat{\tilde{T}}_{\geq \sigma^{(m)}}^{(m)}\right)^{+}\hat{\tilde{T}}_{\geq \sigma^{(m)}}^{(m)}$ spanned by the right singular vectors $\hat{v}_j^{(m)}$ corresponding to singular values greater than the threshold $\sigma^{(m)}$, where $\left(\hat{\tilde{T}}_{\geq \sigma^{(m)}}^{(m)}\right)^{+}$ denotes the Moore-Penrose inverse of $\hat{\tilde{T}}_{\geq \sigma^{(m)}}^{(m)}$. In Step 3, we first add an ancillary register $\ket{0}^{a}$ and then apply a unitary operator $V=\sum_{m=0}^{N-1}V^{(m)}\otimes \ket{m}\bra{m}^{e}$ acting on the register $b$ and $a$ controlled by the register $e$, where $V^{(m)}$ maps $\ket{t}^{b}\ket{0}^{a} \rightarrow \ket{t}^{b}\ket{1}^{a}$ if $t < \sigma^{(m)}$ and $\ket{t}^{b}\ket{0}^{a} \rightarrow \ket{t}^{b}\ket{0}^{a}$ otherwise. Therefore, after Step 3, we get
\begin{align}\label{Step 3}
\ket{\xi_2}&=\frac{1}{||\tilde{\mathcal{T}}(i,:,:)||_F}\sum_{m=0}^{N-1}||\hat{\tilde{\mathcal{T}}}(i,:,m)||_2
\nonumber
\\
&\left(\sum_{j, \overline{\hat{\sigma}}_j^{(m)}\geq \sigma^{(m)}}\beta_j^{(im)}\ket{\hat{v}_j^{(m)}}^{d}\ket{\overline{\hat{\sigma}}_j^{(m)}}^{b}\ket{0}^{a}  \right.
\nonumber
\\
&\left.  \ \ \ \ + \sum_{j, \overline{\hat{\sigma}}_j^{(m)}<\sigma^{(m)}}\beta_j^{(im)}\ket{\hat{v}_j^{(m)}}^{d}\ket{\overline{\hat{\sigma}}_j^{(m)}}^{b}\ket{1}^{a}\right) \ket{m}^{e}.
\end{align}

After the inverse procedure of QSVE in Step 4, (\ref{Step 3}) becomes
\begin{align}\label{Step 4}
&\ket{\xi_3}=\frac{1}{||\tilde{\mathcal{T}}(i,:,:)||_F} \nonumber \\
&\sum_{m=0}^{N-1}||\hat{\tilde{\mathcal{T}}}(i,:,m)||_2\left(\sum_{j, \overline{\hat{\sigma}}_j^{(m)}\geq \sigma^{(m)}}\beta_j^{(im)}\right. \nonumber \\
& \left. \ket{\hat{v}_j^{(m)}}^{d}\ket{0}^{a} + \sum_{j, \overline{\hat{\sigma}}_j^{(m)}<\sigma^{(m)}}\beta_j^{(im)}\ket{\hat{v}_j^{(m)}}^{d}\ket{1}^{a}\right)\ket{m}^{e}.
\end{align}
Then we measure the second register of $\ket{\xi_3}$ and postselect the outcome $\ket{0}^a$ getting
\begin{align} \label{Step 5}
\ket{\xi_4}=\frac{1}{\alpha}\sum_{m=0}^{N-1}\sum_{j,\geq \sigma^{(m)}}\beta_j^{(im)}||\hat{\tilde{\mathcal{T}}}(i,:,m)||_2 \ket{\hat{v}_j^{(m)}}^{d}\ket{m}^{e},
\end{align}
where $$\alpha=\sqrt{\sum \limits_{m=0}^{N-1} \sum \limits_{j,\geq \sigma^{(m)}}||\hat{\tilde{\mathcal{T}}}(i,:,m)||_2^2 \cdot |\beta_j^{(im)}|^2}.$$

Comparing (\ref{eq:25}) with (\ref{Step 5}), we find that the unnormalized state $\sum_{j,\geq \sigma^{(m)}}\beta_j^{(im)}||\hat{\tilde{\mathcal{T}}}(i,:,m)||_2\ket{\hat{v}_j^{(m)}},$ corresponding to the $i$-th row of the truncated matrix $\hat{\tilde{T}}_{\geq \sigma^{(m)}}^{(m)}$, can be seen as an approximation of $\hat{\tilde{\mathcal{T}}}(i,:,m),$ $m=0, \cdots, N-1.$ Hence, $\ket{\xi_4}$ corresponds to an approximation of $\hat{\tilde{\mathcal{T}}}_{\geq \sigma}(i,:,:).$

The probability that we obtain the outcome $\ket{0}$ in Step 5 is
\begin{align}
\frac{||\hat{\tilde{\mathcal{T}}}_{\geq \sigma}(i,:,:)||_F^2}{||\tilde{\mathcal{T}}(i,:,:)||_F^2},
\end{align}
delete this part: where $\hat{\tilde{\mathcal{T}}}_{\geq \sigma}(i,:,:)$ is the $i$-th horizontal slice of the tensor $\hat{\tilde{\mathcal{T}}}_{\geq \sigma}$ whose the $m$-th frontal slice is
$\hat{\tilde{T}}_{\geq \sigma^{(m)}}^{(m)}$. Hence, based on amplitude amplification, we have to repeat the measurement $\mathcal{O}(\frac{||\tilde{\mathcal{T}}(i,:,:)||_F}{||\hat{\tilde{\mathcal{T}}}_{\geq \sigma}(i,:,:)||_F})$ times in order to ensure the success probability of getting the outcome $\ket{0}$ is close to 1.

In Step 6, we perform the inverse QFT on $\ket{\xi_4}$ in (\ref{Step 5}) to get the final state 		
\begin{align} \label{final state 16}
\ket{\xi_5}=&\frac{1}{\alpha \sqrt{N}}\sum_{t,m=0}^{N-1}\sum_{j,\geq \sigma^{(m)}} \beta_j^{(im)}\omega^{-tm} \nonumber \\
&||\hat{\tilde{\mathcal{T}}}(i,:,m)||_2\ket{\hat{v}_j^{(m)}}^{d}\ket{t}^{e},
\end{align}
which corresponds to an approximation of $\tilde{\mathcal{T}}(i,:,:)$, and thus it can also be regarded as an approximation of $\mathcal{T}(i,:,:)$; see the theoretical analysis in Section \ref{Theoretical analysis}.

In the last step, user $i$ is recommended a product $j$ varying with different contexts as needed by measuring the output state $\ket{\xi_5}$. For example, if we need the recommended index at a certain context $t_0$, we can first measure the last register of $\ket{\xi_5}$ in the computational basis and postselect the outcome $\ket{t_0}$, obtaining the state propositional to (unnormalized)
\begin{align}
\sum_{m=0}^{N-1}\sum_{j,\geq \sigma^{(m)}} \beta_j^{(im)}\omega^{-t_0m}||\hat{\tilde{\mathcal{T}}}(i,:,m)||_2\ket{\hat{v}_j^{(m)}}^{d}.
\end{align} We next measure this state in the computational basis to get an index $j$ which is proved to be a good recommendation for user $i$ at context $t_0$.

\subsection{Theoretical analysis} \label{Theoretical analysis}

In this section, the $i$-th horizontal slice of the tensor $\tilde{\mathcal{T}}_{\geq \sigma}$ can be proved to be an approximation of $\mathcal{T}(i,:,:)$. Then sampling from the matrix $\tilde{\mathcal{T}}_{\geq \sigma}(i,:,:)$ yields good recommendations for user $i$; see Theorem \ref{approximation theorem}.
The conclusions of Lemmas \ref{bad recommend} and \ref{threshold and rank} are used in the proof of Theorem \ref{approximation theorem}, so we introduce them first. The proofs of Lemma \ref{threshold and rank} and Theorem \ref{approximation theorem} can be found in Appendices \ref{appendix Lemma 5 } and \ref{appendix Theorem 5 } respectively.

\begin{lemma} \cite{QRS} \label{bad recommend}
	Let $\tilde{A}$ be an approximation of the matrix $A$ such that $||A-\tilde{A}||_F \leq \epsilon||A||_F$. Then, the probability that sampling from $\tilde{A}$ provides a bad recommendation is
	\begin{align}
	\Pr \limits_{(i,j)\sim \tilde{A}} [(i,j) {\rm bad}] \leq \left(\frac{\epsilon}{1-\epsilon}\right)^2.
	\end{align}
\end{lemma}

\blem \label{threshold and rank}
Let $A \in \mathbb{R}^{N \times N}$ be a matrix and $A_k$ be the best rank-$k$ approximation satisfying
$||A-A_k||_F \leq \epsilon||A||_F$. If the threshold for truncating the singular values of $A$ is chosen as $\sigma=\frac{\epsilon||A||_F}{\sqrt{k}}$, then
\begin{align}
||A-A_{\geq \sigma}||_F \leq 2\epsilon||A||_F.
\end{align}
\elem

\bthm \label{approximation theorem}
Algorithm \ref{alg:QRS for tensors} outputs the state $\ket{\tilde{\mathcal{T}}_{\geq \sigma}(i,:,:)}$ corresponding to the approximation of $\mathcal{T}(i,:,:)$ such that for at least $(1-\delta)N$ users, user $i$ in which satisfies
\begin{align}
||\mathcal{T}(i,:,:)-\tilde{\mathcal{T}}_{\geq \sigma}(i,:,:)||_F\leq \epsilon||\mathcal{T}(i,:,:)||_F
\end{align} with probability at least $p_1p_2p_3=(1-e^{-||\mathcal{T}||_F^2/3p})(1-e^{-\zeta^2(\frac{1}{p}-p)\frac{||\mathcal{T}||_F^2}{3N(1+\gamma)}})(1-1/{{\rm poly}N}),$ where $\gamma, \zeta \in [0,1]$ and $p$ is the subsample probability. The precision $\epsilon= \sqrt{(1+\zeta)(\frac{1}{p}-p)} + \epsilon_0 \sqrt{\frac{2(1+\gamma)}{\delta p}},$ $\delta \in (0,1),$ $\epsilon_0=\max \limits_{m=0, \cdots,N-1} 2\epsilon^{(m)}$ if the best rank-$k$ approximation satisfies $||\hat{\tilde{T}}^{(m)}-\hat{\tilde{T}}^{(m)}_{k}||_F \leq \epsilon^{(m)}||\hat{\tilde{T}}^{(m)}||_F$ for a small constant $k$, and the corresponding threshold of each $\hat{\tilde{T}}^{(m)}$ is chosen as $\sigma^{(m)}=\frac{\epsilon^{(m)}||\hat{\tilde{T}}^{(m)}||_F}{\sqrt{k}}.$ Moreover, based on Lemma \ref{bad recommend}, the probability that sampling according to $\tilde{\mathcal{T}}_{\geq \sigma}(i,:,:)$ (is equivalent to measuring the state $\ket{\tilde{\mathcal{T}}_{\geq \sigma}(i,:,:)}$ in the computational basis) provides a bad recommendation is
\begin{align}
\Pr \limits_{t\sim \mathcal{U}_N, j\sim \tilde{\mathcal{T}}_{\geq \sigma}(i,:,:)} [(i,j,t) {\rm bad}] \leq \left(\frac{\epsilon}{1-\epsilon}\right)^2.
\end{align}
\ethm

\subsection{Complexity analysis} \label{complexity of QRS}

The complexity of Algorithm \ref{alg:QRS for tensors}  is given by the following result.

\bthm \label{thm:complexity of qrs}
For at least  $(1-\delta)N$ users, Algorithm \ref{alg:QRS for tensors} outputs an approximation state of $\ket{\mathcal{T}(i,:,:)}$ with complexity $\mathcal{O}(\frac{\sqrt{k}N{\rm polylog}(N) (1+\gamma)}{\min \limits_{m}{\epsilon^{(m)}(1+\epsilon)\sqrt{p}}})$. For suitable parameters,  the running time of Algorithm \ref{alg:QRS for tensors} is $\mathcal{O}(\sqrt{k}N{\rm polylog}(N)).$
\ethm

The proof of Theorem \ref{thm:complexity of qrs} can be found in Appendix  \ref{appendix Theorem 6 }.

Note that the running time of our quantum algorithm depends heavily on the threshold $\sigma^{(m)}=\frac{\epsilon^{(m)}||\hat{\tilde{T}}^{(m)}||_F}{\sqrt{k}}$ which relies on the rank $k$ and corresponding precision $\epsilon^{(m)}$. Above all, the running time of Algorithm \ref{alg:QRS for tensors} is $\mathcal{O}(\sqrt{k}N{\rm polylog}(N) )$ for suitable parameters.
\subsection{A quantum algorithm of tensor completion} \label{Section: another truncate}

In this section, we propose a quantum algorithm for tensor completion based on our quantum t-svd algorithm. This method follows the similar idea of Algorithm \ref{alg:QRS for tensors} but truncate the top $k$ singular values among all the frontal slice $\hat{\tilde{T}}^{(m)}$, $m=0, \cdots, N-1.$ More specifically, in Step 3 of Algorithm \ref{alg:QRS for tensors}, after getting the state $\ket{\xi_1}$, we design another unitary transformation $V'$ acting on the ancillary register $\ket{0}$ that maps $\ket{t}\ket{0} \rightarrow \ket{t}\ket{1}$ if $t < \sigma$ and $\ket{t}\ket{0} \rightarrow \ket{t}\ket{0}$ otherwise, so the state becomes
\begin{align}
&\ket{\xi_2^{'}}= \frac{1}{||\tilde{\mathcal{T}}(i,:,:)||_F}
\nonumber \\
&\left(\sum_{\substack{m,j=0 \\ \overline{\hat{\sigma}}_j^{(m)} \geq \sigma}}^{N-1}||\hat{\tilde{\mathcal{T}}}(i,:,m)||_2 \beta_j^{(im)}\ket{\hat{v}_j^{(m)}}\ket{\overline{\hat{\sigma}}_j^{(m)}}\ket{0}\right. \nonumber \\
& \left. + \sum_{\substack{m,j=0 \\ \overline{\hat{\sigma}}_j^{(m)} < \sigma}}^{N-1}||\hat{\tilde{\mathcal{T}}}(i,:,m)||_2\beta_j^{(im)}\ket{\hat{v}_j^{(m)}}\ket{\overline{\hat{\sigma}}_j^{(m)}}\ket{1}\right)\ket{m}.
\end{align}

Then after the inverse QSVE, measuring the third register and postselecting the outcome $\ket{0}$, just as done in Steps 4 and 5 of Algorithm \ref{alg:QRS for tensors}, we get
\begin{align}
\ket{\xi_4^{'}}=\frac{1}{\alpha^{'}}\sum_{\substack{m,j=0 \\ \overline{\hat{\sigma}}_j^{(m)} \geq \sigma}}^{N-1}||\hat{\tilde{\mathcal{T}}}(i,:,m)||_2 \beta_j^{(im)}\ket{\hat{v}_j^{(m)}}\ket{m},
\end{align}
where $\alpha^{'}=\left(\sum \limits_{\substack{m,j=0 \\ \overline{\hat{\sigma}}_j^{(m)} \geq \sigma}}^{N-1}||\hat{\tilde{\mathcal{T}}}(i,:,m)||_2^2 \cdot |\beta_j^{(im)}|^2\right)^{1/2}.$

The last step is the inverse QFT  which outputs the final state
\begin{align}
&\ket{\xi_5^{'}}= \nonumber \\
&\frac{1}{\alpha^{'}\sqrt{N}}\sum_{t=0}^{N-1}\sum_{\substack{m,j=0 \\ \overline{\hat{\sigma}}_j^{(m)} \geq \sigma}}^{N-1}||\hat{\tilde{\mathcal{T}}}(i,:,m)||_2 \beta_j^{(im)}\omega^{-mt}\ket{\hat{v}_j^{(m)}}\ket{t}.
\end{align}

Our first truncation method applied in quantum recommendation systems introduced in Section \ref{Section QRS for tensors} is called t-svd-tubal compression and the second algorithm in Section \ref{Section: another truncate} is called t-svd compression. According to the comparison and analysis of these two methods in \cite{Novel}, although the latter has better performance when applied to stationary camera videos, the former works much better on the non-stationary panning camera videos because it better captures the convolution relations between different frontal slices of the tensor in dynamic video, so we design the quantum version of both methods in this paper.

\section{Conclusion}

The main contribution of this paper consists of two parts. First, we present a quantum t-svd algorithm for third-order tensors which achieves the complexity of $\mathcal{O}(N{\rm polylog}(N))$. The other innovation is that we propose the first quantum algorithm for recommendation systems modeled by third-order tensors. We prove that our algorithm can provide good recommendations varying with contexts and run in expected time $\mathcal{O}(N{\rm polylog}(N){\rm poly}(k))$ for some suitable parameters, which is exponentially faster than known classical algorithms. We also propose a variant of Algorithm \ref{alg:QRS for tensors}, which deals with third-order tensors completion problems. 


\appendix

\section{The proof of Theorem \ref{the QSVE for tensor Ahat}} \label{appendix Theorem 3.1 }

In this appendix, we prove Theorem \ref{the QSVE for tensor Ahat}.

In the QSVE algorithm \cite{QRS}, Kerenidis and Prakash first constructed two isometries $P$ and $Q$ which are implemented efficiently through two unitary transformations $U_P$ and $U_Q$, such that the target matrix $A$ has the factorization $\frac{A}{||A||_F}=P^{\dagger}Q$. Based on these two isometries, the unitary operator $W=(2PP^{\dagger}-I_{mn})(2QQ^{\dagger}-I_{mn})$ can be implemented efficiently. The QSVE algorithm utilizes the connection between the eigenvalues $e^{\pm i \theta_i}$ of $W$ and the singular values $\sigma_i$ of $A$, i.e., $\cos \frac{\theta_i}{2}=\frac{\sigma_i}{||A||_F}.$ Therefore, we can perform the phase estimation on $W$ to get an estimated value $ \overline{\theta}_i$ and then compute the estimated singular value $\overline{\sigma}_{i}$ stored in a register superposed with its corresponding singular vector.

In the proof of Theorem \ref{the QSVE for tensor Ahat}, we assume that every frontal slice of the tensor $\mathcal{A}$ is stored in the data structure stated in Lemma \ref{data structure}.  Then according to Theorem 5.1 in \cite{QRS}, the quantum state $\ket{A_i^{(k)}}$ and $\ket{\boldsymbol{s}_A^{(k)}}$ can be prepared efficiently by the operators $P^{(k)}$ and $Q^{(k)}$, $k=0,\cdots,N_3-1$. Based on our quantum t-svd algorithm, the QSVE is expected to be performed on each frontal slice of $\hat{\mathcal{A}}$, denoted as $\hat{A}^{(m)}=\frac{1}{\sqrt N_3}\sum_{k=0}^{N_3-1}\omega^{km}A^{(k)}$. For achieving this, we construct two isometries $\hat{P}^{(m)}$ and $\hat{Q}^{(m)}$. According to Remark \ref{modified QSVE remarks}, the input is chosen as the state $\ket{\hat{A}^{(m)}}=\frac{1}{||\hat{A}^{(m)}||_F}\sum_{i,j,k}\omega^{km}\mathcal{A}(i,j,k)\ket{i}\ket{j}$. Following the similar procedure of the QSVE algorithm \cite{QRS}, we can obtain the desired output state.
\begin{proof}
	Since every $A^{(k)}$, $k=0,\cdots,N_3-1$, is stored in the binary tree structure, the quantum computer can perform the following mappings in $O({\rm polylog}(N_1N_2))$ time, as shown in Theorem 5.1 in \cite{QRS}:
	\begin{align}
	&U_{P}^{(k)}: \ket{i}\ket{0}
	\nonumber \\
	&\hspace{3ex}\rightarrow \ket{i}\ket{A_i^{(k)}}=\frac{1}{||A_i^{(k)}||_2}\sum_{j=0}^{N_2-1}\mathcal{A}(i,j,k)\ket{i}\ket{j},
	\nonumber \\
	& U_{Q}^{(k)}: \ket{0}\ket{j}
	\nonumber \\
	&\hspace{3ex}\rightarrow \ket{\boldsymbol{s}_A^{(k)}}\ket{j}=\frac{1}{||A^{(k)}||_F}\sum_{i=0}^{N_1-1}||A_i^{(k)}||_2\ket{i}\ket{j},
	\end{align}
	where $A_i^{(k)}$ is the $i$-th row of $A^{(k)}$ and $\boldsymbol{s}_A^{(k)}\triangleq \frac{1}{||A||_F}\left[||A_0^{(k)}||_2 , ||A_1^{(k)}||_2 , \cdots , ||A_{N_1-1}^{(k)}||_2 \right]^{T},$ $k=0,\cdots,N_3-1.$
	
	We can define two isometries $P^{(k)} \in \mathbb{R}^{N_1N_2 \times N_1}$ and $Q^{(k)}\in \mathbb{R}^{N_1N_2 \times N_2}$  related to $U_{P}^{(k)}$ and $U_{Q}^{(k)}$ as followed:
	\begin{align} \label{PQ}
	P^{(k)}=\sum_{i=0}^{N_1-1}\ket{i}\ket{A_i^{(k)}}\bra{i} , \quad Q^{(k)}=\sum_{j=0}^{N_1-1}\ket{\boldsymbol{s}_A^{(k)}}\ket{j}\bra{j}.
	\end{align}
	
	Define another operator $\hat{P}^{(m)}\triangleq \frac{1}{\sqrt{N_3}}\sum_{k=0}^{N_3-1}\frac{||A_i^{(k)}||_2\omega^{km}}{||\hat{A}_i^{(m)}||_2}P^{(k)}$ which achieves the state preparation of the rows of the matrix $\hat{A}^{(m)}$. Since every isometry $P^{(k)}$, $k=0,\cdots, N_3-1,$ can be implemented with complexity $\mathcal{O}(\log N_1)$, $\hat{P}^{(m)}$ can also be implemented efficiently. Substituting $P^{(k)}$ into $\hat{P}^{(m)}$, we have for each $m=0, \cdots, N_3-1,$
	\begin{align}
	\hat{P}^{(m)}&=\frac{1}{\sqrt{N_3}}\sum_{k=0}^{N_3-1}\frac{||A_i^{(k)}||_2\omega^{km}}{||\hat{A}_i^{(m)}||_2}\sum_{i}\ket{i}\ket{A_i^{(k)}}\bra{i}\nonumber \\
	&=\sum_{i}\ket{i}\ket{\hat{A}_i^{(m)}}\bra{i},
	\end{align} where $$\ket{\hat{A}_i^{(m)}}=\ket{\mathcal{\hat{A}}(i,:,m)}=\frac{1}{\sqrt{N_3}}\sum_{k=0}^{N_3-1}\frac{||A_i^{(k)}||_2\omega^{km}}{||\hat{A}_i^{(m)}||_2}\ket{A_i^{(k)}}.$$ It is easy to check that $\hat{P}^{(m)}$ is an isometry:
	\begin{align}
	\hat{P}^{^{(m)}\dagger}\hat{P}^{(m)}&=(\sum_{i}\ket{i}\bra{i}\bra{\hat{A}_i^{(m)}})(\sum_{j}\ket{j}\ket{\hat{A}_j^{(m)}}\bra{j})\nonumber\\
&=I_{N_1}.
	\end{align}
	
	We construct another array of $N_3$ binary trees, each of which has the root storing $||\hat{A}^{(m)}||_F^2$ and the $i$-th leaf storing $||\hat{A}_i^{(m)}||_2^2$. Define $\hat{\boldsymbol{s}}_A^{(m)}=\frac{1}{||\hat{A}^{(m)}||_F}\begin{bmatrix}
	||\hat{A}_0^{(m)}||_2 & ||\hat{A}_1^{(m)}||_2 & \cdots & ||\hat{A}_{N_1-1}^{(m)}||_2
	\end{bmatrix}^{T}$. According to the proof of Lemma 5.3 in \cite{QRS}, we can perform the mapping
	\begin{align} \label{UQhatm}
	\hat{U}_{Q}^{(m)}: \ket{0}\ket{j}\rightarrow &\ket{\hat{\boldsymbol{s}}_A^{(m)}}\ket{j} \nonumber \\
	=&\frac{1}{||\hat{A}^{(m)}||_F}\sum_{i}||\hat{A}_i^{(m)}||_2\ket{i}\ket{j}
	\end{align} and the corresponding isometry $\hat{Q}^{(m)}=\sum_{j}\ket{\hat{\boldsymbol{s}}_A^{(m)}}\ket{j}\bra{j}$ satisfies $\hat{Q}^{^{(m)}\dagger}\hat{Q}^{(m)}=I_{N_2}$.
	
	Now we can perform QSVE on the matrix $\hat{A}^{(m)}$. First, the factorization $\frac{\hat{A}^{(m)}}{||\hat{A}^{(m)}||_F}=\hat{P}^{(m)\dagger}\hat{Q}^{(m)}$ can be easily verified.
	Moreover, we can prove that $2\hat{P}^{(m)}\hat{P  }^{^{(m)}\dagger}-I_{N_1N_2}$ is unitary and it can be efficiently implemented in time $\mathcal{O}({\rm polylog}(N_1N_2))$. Actually,
	\begin{align}
	&2\hat{P}^{(m)}\hat{P  }^{^{(m)}\dagger}-I_{N_1N_2} \nonumber \\
	=&2\sum_{i}\ket{i}\ket{\hat{A}_i^{(m)}}\bra{i}\bra{\hat{A}_i^{(m)}}-I_{N_1N_2} \nonumber \\
	=&U_{\hat{P}^{(m)}}\left[2\sum_{i}\ket{i}\ket{0}\bra{i}\bra{0}-I_{N_1N_2} \right]U_{\hat{P}^{(m)}}^{\dagger},
	\end{align}
	where $2\sum_{i}\ket{i}\ket{0}\bra{i}\bra{0}-I_{N_1N_2}$ is a reflection. $U_{\hat{P}^{(m)}}$ is the unitary operator corresponding to the isometry $\hat{P}^{(m)}$, i.e., $U_{\hat{P}^{(m)}}=\sum_{i}\ket{i}\ket{\hat{A}_i^{(m)}}\bra{i}\bra{0}$. The similar result holds for $2\hat{Q}^{(m)}\hat{Q}^{^{(m)}\dagger}-I_{N_1N_2}$.
	
	Now denote
	\begin{align} \label{W_m}
	W_m=(2\hat{P}^{(m)}\hat{P}^{^{(m)}\dagger}-I_{N_1N_2})(2\hat{Q}^{(m)}\hat{Q}^{^{(m)}\dagger}-I_{N_1N_2}),
	\end{align} and we can prove that the subspace spanned by $\{\hat{Q}^{(m)}\ket{\hat{v}_i^{(m)}}, \hat{P}^{(m)}\ket{\hat{u}_i^{(m)}}\}$ is invariant under the unitary transformation $W_m$:
	\begin{align*}
	W_m\hat{Q}^{(m)}\ket{\hat{v}_i^{(m)}}
	&=\frac{2\hat{\sigma}_i^{(m)}}{||\hat{A}^{(m)}||_F}\hat{P}^{(m)}\ket{\hat{u}_i^{(m)}}-Q\ket{\hat{v}_i^{(m)}},
	\end{align*}
	\begin{align*}
	&W_m\hat{P}^{(m)}\ket{\hat{u}_i^{(m)}}= \nonumber \\
	&\left( \frac{4\hat{\sigma}_{i}^2}{||\hat{A}^{(m)}||_F}-1 \right)\hat{P}^{(m)}\ket{\hat{u}_i^{(m)}}-\frac{2\hat{\sigma}_{i}}{||\hat{A}^{(m)}||_F}\hat{Q}^{(m)}\ket{\hat{v}_i^{(m)}}.
	\end{align*}
	The matrix $W_m$ can be calculated under an orthonormal basis using the Schmidt orthogonalization,
	and it is a rotation in the subspace spanned by its eigenvectors $\ket{\omega^{(m)}_{i \pm}}$ corresponding to eigenvalues $e^{\pm i\theta_i^{(m)}}$, where $\theta_i^{(m)}$ is the rotation angle satisfying $\cos(\theta_i^{(m)}/2)=\frac{\hat{\sigma}_i^{(m)}}{||\hat{A}^{(m)}||_F}$, i.e.
	\begin{align*}
	&\hat{Q}^{(m)}\ket{\hat{v}_i^{(m)}}=\sqrt{2}(\ket{\omega^{(m)}_{i +}}+\ket{\omega^{(m)}_{i -}}) \\
	&\hat{P}^{(m)}\ket{\hat{u}_i^{(m)}}=\sqrt{2}(e^{ i\theta_i/2}\ket{\omega^{(m)}_{i +}}+e^{ -i\theta_i/2}\ket{\omega^{(m)}_{i -}}).
	\end{align*}
	
	In the QSVE algorithm on the matrix $\hat{A}^{(m)}$, $m=0, \cdots, N_3-1,$ we choose the input state as the Kronecker product form of the normalized matrix $\frac{\hat{A}^{(m)}}{||\hat{A}^{(m)}||_F}$ represented in the svd, i.e.,  $\ket{\hat{A}^{(m)}} =\frac{1}{||\hat{A}^{(m)}||_F}\sum_{i}\hat{\sigma}_i^{(m)}\ket{\hat{u}_i^{(m)}} \ket{\hat{v}_i^{(m)}}$. Then
	\begin{align}
	&I_{N_1}\otimes\hat{U}_{Q^{(m)}}\ket{\hat{A}^{(m)}}
	\nonumber \\
	=&\frac{1}{||\hat{A}^{(m)}||_F}\sum_{i}\sqrt{2}\hat{\sigma}_i^{(m)}\ket{\hat{u}_i^{(m)}}(\ket{\omega^{(m)}_{i +}}+\ket{\omega^{(m)}_{i -}}).
	\end{align}
	Performing the phase estimation on $W_m$ and computing the estimated singular value of $\hat{A}^{(m)}$ through oracle $\hat{\sigma}_{i}^{(m)}=||\hat{A}^{(m)}||_F\cos(\theta^{(m)}_i/2),$ we obtain
	\begin{align}
		\frac{1}{||\hat{A}^{(m)}||_F}\sum_{i}\sqrt{2}&\hat{\sigma}_{i}^{(m)}\ket{\hat{u}_i^{(m)}}\left(\ket{\omega^{(m)}_{i +}}\ket{\overline{\theta}_i^{(m)}}+ \right. \nonumber \\
		&\left. \ket{\omega^{(m)}_{i -}}\ket{-\overline{\theta}_i^{(m)}}\right)
		\ket{\overline{\sigma}_{i}^{(m)}}.
	\end{align}
	we next uncompute the phase estimation procedure and then apply the inverse of $I_{N_1}\otimes\hat{U}_{Q^{(m)}}$, obtaining the desired state \eqref{eq:1 feb3} in Theorem \ref{the QSVE for tensor Ahat}..
\end{proof}

\section{The proof of Lemma \ref{threshold and rank}} \label{appendix Lemma 5 }
\begin{proof}
	Let $\sigma_i$ denote the singular value of $A$ and $\ell$ be the largest integer for which $\sigma_{\ell} \geq \frac{\epsilon||A||_F}{\sqrt{k}}.$ By the triangle inequality, $||A-A_{\geq \sigma}||_F \leq ||A-A_k||_F +||A_k-A_{\geq \sigma}||_F.$
	If $k \leq \ell$, it's easy to conclude that $||A_k-A_{\geq \sigma}||_F \leq ||A-A_k||_F \leq \epsilon||A||_F$.
	If $k > \ell$, $||A_k-A_{\geq \sigma}||_F^2=\sum_{i=\ell+1}^{k}\sigma_i^2 \leq k\sigma_{\ell+1}^2 \leq k(\frac{\epsilon||A||_F}{\sqrt{k}})^2 \leq (\epsilon||A||_F)^2$. Above all, we have $||A-A_{\geq \sigma}||_F \leq 2\epsilon||A||_F$.
\end{proof}

\section{The proof of Theorem \ref{approximation theorem}} \label{appendix Theorem 5 }
\begin{proof}
	
	Based on Lemma \ref{threshold and rank} in the main text, if the best rank-$k$ approximation satisfies $||\hat{\tilde{T}}^{(m)}-\hat{\tilde{T}}^{(m)}_{k}||_F \leq
	\epsilon^{(m)}||\hat{\tilde{T}}^{(m)}||_F, $ then
	\begin{align} \label{eq40}
	||\hat{\tilde{T}}^{(m)}-\hat{\tilde{T}}^{(m)}_{\geq \sigma^{(m)}}||_F \leq 2
	\epsilon^{(m)}||\hat{\tilde{T}}^{(m)}||_F \leq \epsilon_0||\hat{\tilde{T}}^{(m)}||_F,
	\end{align} for $\sigma^{(m)}=\frac{\epsilon^{(m)}||\hat{\tilde{T}}^{(m)}||_F}{\sqrt{k}}, m=0, \cdots, N-1.$ By summarizing on both side of (\ref{eq40}), we get
	\begin{align} \label{tensor inequality}
	||\hat{\tilde{\mathcal{T}}}-\hat{\tilde{\mathcal{T}}}_{\geq \sigma}||_F^2 = \sum_{m=0}^{N-1}||\hat{\tilde{T}}^{(m)}-\hat{\tilde{T}}^{(m)}_{\geq \sigma^{(m)}}||_F^2 \leq \epsilon_0^2||\hat{\tilde{\mathcal{T}}}||_F^2.
	\end{align}
	Since the inverse QFT along the third mode of the tensor $\mathcal{T}$ cannot change the Frobenius norm of its horizontal slice, \eqref{tensor inequality} can be be re-written as
	\begin{align}
||\tilde{\mathcal{T}}-\tilde{\mathcal{T}}_{\geq \sigma}||_F^2 \leq \epsilon_0^2||\tilde{\mathcal{T}}||_F^2.
	\end{align}
Moreover, noticing that $||\tilde{\mathcal{T}}-\tilde{\mathcal{T}}_{\geq \sigma}||_F^2=\sum_{i=0}^{N-1}||\tilde{\mathcal{T}}(i,:,:)-\tilde{\mathcal{T}}_{\geq \sigma}(i,:,:)||_F^2$, we have	 $\E \left(||\tilde{\mathcal{T}}(i,:,:)-\tilde{\mathcal{T}}_{\geq \sigma}(i,:,:)||_F^2\right) \leq \frac{\epsilon_0^2||\tilde{\mathcal{T}}||_F^2}{N}.$
	Due to Markov's Inequality (\cite[Proposition 2.6]{R06}),
	\begin{align}
	&\Pr \left(||\tilde{\mathcal{T}}(i,:,:)-\tilde{\mathcal{T}}_{\geq \sigma}(i,:,:)||_F^2 > \frac{\epsilon_0^2||\tilde{\mathcal{T}}||_F^2}{\delta N}\right) \nonumber \\
	 \leq &\frac{\E \left(||\tilde{\mathcal{T}}(i,:,:)-\tilde{\mathcal{T}}_{\geq \sigma}(i,:,:)||_F^2\right)\delta N}{\epsilon_0^2||\tilde{\mathcal{T}}||_F^2} = \delta
	\end{align} holds for some $\delta \in (0,1) $. That means at least $(1-\delta)N$ users $i$ satisfy
	\begin{align} \label{eq111}
	||\tilde{\mathcal{T}}(i,:,:)-\tilde{\mathcal{T}}_{\geq \sigma}(i,:,:)||_F^2 \leq \frac{\epsilon_0^2||\tilde{\mathcal{T}}||_F^2}{\delta N}.
	\end{align}
	
	During the preprocessing part of Algorithm \ref{alg:QRS for tensors}, tensor $\tilde{\mathcal{T}}$ is obtained by sampling the tensor $\mathcal{T}$ with uniform probability $p$, so $\E \left(||\tilde{\mathcal{T}}||_F^2\right)=||\mathcal{T}||_F^2/p$. Using the Chernoff bound, we have $\Pr \left( ||\tilde{\mathcal{T}}||_F^2 >(1+\theta)||\mathcal{T}||_F^2/p\right) \leq e^{-\theta^2||\mathcal{T}||_F^2/3p}$ for $\theta \in [0,1]$, which is exponentially small. Here, we choose $\theta=1$, then the probability that
	\begin{align}
	||\tilde{\mathcal{T}}||_F^2 \leq 2||\mathcal{T}||_F^2/p
	\end{align} is $p_1=1-e^{-||\mathcal{T}||_F^2/3p}$.  \label{eq 45}
	Based on the third assumption in Assumption \ref{assumption for QRS}, we sum both sides of (\ref{123}) for $m$ and $i$ respectively, obtaining
	\begin{align}
	\frac{1}{1+\gamma}\frac{||\mathcal{T}||_F^2}{N} \leq ||\mathcal{T}(i,:,:)||_F^2 \leq (1+\gamma)\frac{||\mathcal{T}||_F^2}{N},
	\end{align} and
	\begin{align}
	\frac{1}{1+\gamma}\frac{||\mathcal{T}||_F^2}{N} \leq ||T^{(m)}||_F^2 \leq (1+\gamma)\frac{||\mathcal{T}||_F^2}{N}.
	\end{align}
	Then,  (\ref{eq111}) becomes
	\begin{align} \label{eq114}
	||\tilde{\mathcal{T}}(i,:,:)-\tilde{\mathcal{T}}_{\geq \sigma}(i,:,:)||_F^2
	\leq \frac{2\epsilon_0^2(1+\gamma)}{\delta p}||\mathcal{T}(i,:,:)||_F^2
	\end{align} with probability $p_1$.
	
	Meanwhile, since
	\begin{align*}
	\E \left(||\mathcal{T}(i,:,:)-\tilde{\mathcal{T}}(i,:,:)||_F^2\right) =(\frac{1}{p}-p)||\mathcal{T}(i,:,:)||_F^2,
	\end{align*}
	then
 \begin{align}
& \Pr \left(||\mathcal{T}(i,:,:)-\tilde{\mathcal{T}}(i,:,:)||_F^2 > \nu ||\mathcal{T}(i,:,:)||_F^2\right) \nonumber \\
& \leq e^{-\zeta^2(\frac{1}{p}-p)\frac{||\mathcal{T}||_F^2}{3N(1+\gamma)}},
 \end{align}
where $\nu=(1+\zeta)(\frac{1}{p}-p)$ and $\zeta \in [0,1]$. That means with probability at least $p_2=1-e^{-\zeta^2(\frac{1}{p}-p)\frac{||\mathcal{T}||_F^2}{3N(1+\gamma)}}$,
	\begin{align} \label{eq113}
	||\mathcal{T}(i,:,:)-\tilde{\mathcal{T}}(i,:,:)||_F^2 \leq \nu||\mathcal{T}(i,:,:)||_F^2.
	\end{align}
	Combining (\ref{eq114}) and (\ref{eq113}) together and by triangle inequality, we obtain
	\begin{align} \label{eq115}
	& ||\mathcal{T}(i,:,:)-\tilde{\mathcal{T}}_{\geq \sigma}(i,:,:)||_F \nonumber \\
    \leq &||\mathcal{T}(i,:,:)-\tilde{\mathcal{T}}(i,:,:)||_F +||\tilde{\mathcal{T}}(i,:,:)-\tilde{\mathcal{T}}_{\geq \sigma}(i,:,:)||_F \nonumber \\
	\leq &\epsilon ||\mathcal{T}(i,:,:)||_F.
	\end{align}
	According to Lemma \ref{bad recommend}, the probability that sampling according to $\tilde{\mathcal{T}}_{\geq \sigma}(i,:,:)$ provides a bad recommendation is
	\begin{align}
	\Pr \limits_{t\sim \mathcal{U}_N, j\sim \tilde{\mathcal{T}}_{\geq \sigma}(i,:,:)} [(i,j,t) {\rm bad}] \leq \left(\frac{\epsilon}{1-\epsilon}\right)^2.
	\end{align}
\end{proof}

\section{The proof of Theorem \ref{thm:complexity of qrs}} \label{appendix Theorem 6 }
\begin{proof}
	Similar to the complexity of Algorithm \ref{alg:Quantum t-svd}, the QFT is performed with the complexity $\mathcal{O}((\log {N})^2)$. The QSVE algorithm takes time $\mathcal{O}(N{\rm polylog}(N)/{\epsilon_{\rm SVE}^{(m)}})$ and outputs the superposition state with probability $p_3=1-1/{{\rm poly}N}.$
	
	In Step 5, we need to repeat the measurement $\mathcal{O}\left(\frac{||\tilde{\mathcal{T}}(i,:,:)||_F}{||\hat{\tilde{\mathcal{T}}}_{\geq \sigma}(i,:,:)||_F}\right)$ times in order to ensure the probability of getting the outcome $\ket{0}$ in Step 5 is close to 1. For most users, we can prove that $\frac{||\tilde{\mathcal{T}}(i,:,:)||_F}{||\hat{\tilde{\mathcal{T}}}_{\geq \sigma}(i,:,:)||_F}$ is bounded and the upper bound is a constant for appropriate parameters. The proof is in the following.
	
	Since $\E \left(||\tilde{\mathcal{T}}(i,:,:)||_F^2\right)=\frac{||\mathcal{T}(i,:,:)||_F^2}{p} \leq (1+\gamma)\frac{||\mathcal{T}||_F^2}{pN},$ then by Chernoff bound,
	\begin{align} \label{eq117}
	||\tilde{\mathcal{T}}(i,:,:)||_F^2 \leq \frac{2(1+\gamma)||\mathcal{T}||_F^2}{pN}
	\end{align} holds with probability close to 1.	
	Moreover, from the previous discussion, there are at least $(1-\delta)N$ users satisfying  $||\mathcal{T}(i,:,:)-\tilde{\mathcal{T}}_{\geq \sigma}(i,:,:)||_F\leq \epsilon ||\mathcal{T}(i,:,:)||_F$, then
	$(1+\epsilon)||\mathcal{T}(i,:,:)||_F \leq ||\tilde{\mathcal{T}}_{\geq \sigma}(i,:,:)||_F \leq (1+\epsilon)||\mathcal{T}(i,:,:)||_F.$ Since the Frobenius norm is unchanged under the Fourier transform, we get
	\begin{align}
	(1+\epsilon)||\hat{T}_{(i)}||_F \leq ||\hat{\tilde{\mathcal{T}}}_{\geq \sigma}(i,:,:)||_F \leq (1+\epsilon)||\hat{T}_{(i)}||_F.
	\end{align}
Therefore,
	\begin{align} \label{eq118}
	||\hat{\tilde{\mathcal{T}}}_{\geq \sigma}(i,:,:)||_F^2 \geq (1+\epsilon)^2||\hat{T}_{(i)}||_F^2\geq \frac{(1+\epsilon)^2}{1+\gamma}\frac{||\mathcal{T}||_F^2}{N}.
	\end{align}
	
	Combining (\ref{eq117}) and (\ref{eq118}) together, we can conclude that for at least $(1-\delta)N$ users, $\frac{||\tilde{\mathcal{T}}(i,:,:)||_F}{||\hat{\tilde{\mathcal{T}}}_{\geq \sigma}(i,:,:)||_F}$ is bounded with probability $p_1p_2$, that is,
	\begin{align} \label{eq55}
	\frac{||\tilde{\mathcal{T}}(i,:,:)||_F}{||\hat{\tilde{\mathcal{T}}}_{\geq \sigma}(i,:,:)||_F} \leq \left(\frac{(1+\gamma)\frac{2||\mathcal{T}||_F^2}{pN}}{\frac{(1+\epsilon)^2}{1+\gamma}\frac{||\mathcal{T}||_F^2}{N}}\right)^{1/2} = \frac{\sqrt{2}(1+\gamma)}{(1+\epsilon)\sqrt{p}}.
	\end{align}

	The precision for the singular value estimation algorithm on the matrix $||\hat{\tilde{T}}^{(m)}||_F$ can be chosen as $\epsilon_{\rm SVE}^{(m)}=\frac{\sigma^{(m)}}{||\hat{\tilde{T}}^{(m)}||_F}.$ Therefore, the total complexity of Algorithm \ref{alg:QRS for tensors} is
	\begin{align*}
	&({{\rm log}N})^4 \cdot \frac{N{\rm polylog}(N)}{\min \limits_{m}{\epsilon_{\rm SVE}^{(m)}}} \cdot \frac{||\tilde{\mathcal{T}}(i,:,:)||_F}{||\hat{\tilde{\mathcal{T}}}_{\geq \sigma}(i,:,:)||_F} \\
	 \leq & ({{\rm log}N})^4{N\rm polylog}N \max \limits_{m}{\frac{||\hat{\tilde{T}}^{(m)}||_F}{\sigma^{(m)}}}\cdot \frac{\sqrt{2}(1+\gamma)}{(1+\epsilon)\sqrt{p}} \\
	 \approxeq & \frac{\sqrt{k}N{\rm polylog}(N) (1+\gamma)}{\min \limits_{m}{\epsilon^{(m)}(1+\epsilon)\sqrt{p}}},
	\end{align*}
	where $\epsilon^{(m)}$ and $\epsilon$ are defined individually in Theorem \ref{approximation theorem}.	
\end{proof}

\nocite{*}
\bibliographystyle{plain}
\bibliography{bibliography_arxiv}

\end{document}